\documentclass[]{interact}

\usepackage{epstopdf}
\usepackage{subfigure}

\usepackage{amsmath}

\usepackage{hyperref}
\hypersetup{
  colorlinks = true,
  citecolor = blue,
  urlcolor = blue
}

\usepackage{braket}

\newcommand{{\bfC}}{\mbox{\boldmath$C$\unboldmath}}
\newcommand{{\bfu}}{\mbox{\boldmath$u$\unboldmath}}
\newcommand{{\bfp}}{\mbox{\boldmath$p$\unboldmath}}
\newcommand{{\bfr}}{\mbox{\boldmath$r$\unboldmath}}
\newcommand{{\bfv}}{\mbox{\boldmath$v$\unboldmath}}
\newcommand{{\bff}}{\mbox{\boldmath$f$\unboldmath}}
\newcommand{{\bfF}}{\mbox{\boldmath$F$\unboldmath}}
\newcommand{{\bfA}}{\mbox{\boldmath$A$\unboldmath}}
\newcommand{{\bfmu}}{\mbox{\boldmath$\mu$\unboldmath}}
\newcommand{{\bfchi}}{\mbox{\boldmath$\chi$\unboldmath}}
\newcommand{{\bfphi}}{\mbox{\boldmath$\phi$\unboldmath}}
\newcommand{{\bflambda}}{\mbox{\boldmath$\lambda$\unboldmath}}
\newcommand{\gradv}{\boldsymbol{\nabla}}

\newcommand{{\cL}}{\mbox{\boldmath${\cal L}$\unboldmath}}
\newcommand{{\cJ}}{\mbox{\boldmath${\cal J}$\unboldmath}}
\newcommand{{\cS}}{\mbox{\boldmath${\cal S}$\unboldmath}}

\newcommand{{\cF}}{\mbox{\boldmath${\cal F}$\unboldmath}}
\newcommand{{\cG}}{\mbox{\boldmath${\cal G}$\unboldmath}}
\newcommand{{\cE}}{\mbox{\boldmath${\cal E}$\unboldmath}}
\newcommand{{\cB}}{\mbox{\boldmath${\cal B}$\unboldmath}}
\newcommand{{\cX}}{\mbox{\boldmath${\cal X}$\unboldmath}}
\newcommand{{\cY}}{\mbox{\boldmath${\cal Y}$\unboldmath}}
\newcommand{{\cM}}{\mbox{\boldmath${\cal M}$\unboldmath}}
\def\v#1{{\bf#1}}

\DeclareRobustCommand{\uvec}[1]{{%
  \ifcat\relax\noexpand#1%
    \bm{\hat{#1}}%
  \else
    \ifcsname uvec#1\endcsname
      \csname uvec#1\endcsname
    \else
      \bm{\hat{\mathbf{#1}}}%
     \fi
   \fi
}}

\usepackage[numbers,sort&compress]{natbib}
\bibpunct[, ]{[}{]}{,}{n}{,}{,}
\makeatletter
\def\NAT@def@citea{\def\@citea{\NAT@separator}}
\makeatother

\begin{document}

\title{\Large Dirac quantisation condition: a comprehensive review}

\author{ \name{Ricardo Heras\thanks{Email:
      ricardo.heras.13@ucl.ac.uk}} \affil{Department of Physics and Astronomy,\\
 University College London, London, WC1E 6BT, UK} }

\maketitle

\begin{abstract}
In most introductory courses on electrodynamics, one is taught the electric charge is quantised but no theoretical explanation related
to this law of nature is offered. Such an explanation is postponed to graduate courses on electrodynamics, quantum mechanics and quantum field theory, where the famous Dirac quantisation condition is introduced, which states that a single magnetic monopole in the Universe would explain the electric charge quantisation. Even when this condition assumes the existence of a not-yet-detected magnetic monopole, it provides the most accepted explanation for the observed quantisation of the electric charge. However, the usual derivation of the Dirac quantisation condition involves the subtle concept of an ``unobservable'' semi-infinite magnetised line, the so-called ``Dirac string,'' which may be difficult to grasp in a first view of the subject. The purpose of this review is to survey the concepts underlying the Dirac quantisation condition, in a way that may be accessible to advanced undergraduate and graduate students. Some of the discussed concepts are gauge invariance, singular potentials, single-valuedness of the wave function, undetectability of the Dirac string and quantisation of the electromagnetic angular momentum. Five quantum-mechanical and three semi-classical derivations of the Dirac quantisation condition are reviewed. In addition, a simple derivation of this condition involving heuristic and formal arguments is presented.
\end{abstract}

\tableofcontents

\section{Introduction}
\label{1}

\begin{figure}[h]
  \centering
  \includegraphics[width=208pt]{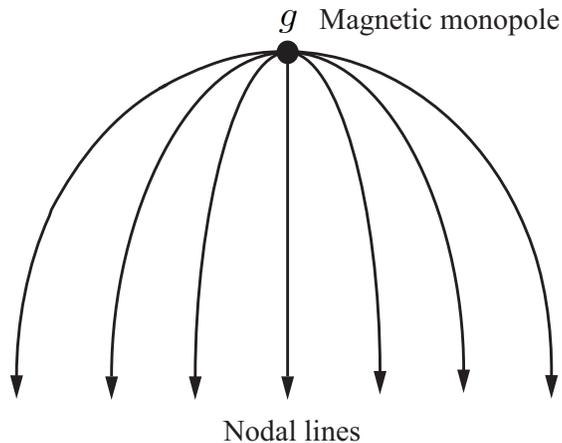}
  \caption{Nodal lines as envisioned by Dirac.}
  \label{Fig1}
\end{figure}

\noindent In the early months of 1931, Dirac was seeking for an explanation of the observed fact that the electric charge is always quantised \cite{1}.
In his quest for explaining this mysterious charge quantisation, he incidentally came across with the idea of  magnetic monopoles, which turned out to be of vital importance for his ingenious explanation presented in his 1931 paper \cite{2}. In this seminal paper, Dirac  envisioned hypothetical nodal lines to be semi-infinite magnetised lines with vanishing wave function and having the same end point, which is the singularity of the magnetic field where the monopole is located (see Figure~\ref{Fig1}). A quantum-mechanical argument on these nodal lines led him to his celebrated quantisation condition: $q g=n\hbar c/2$. Here, $q$ and $g$ denote electric and magnetic charges,  $\hbar$ is the reduced Planck's constant, $c$ is the speed of light, $n$ represents an integer number, and
 we are adopting Gaussian units. Dirac wrote \cite{2}: ``Thus at the end point [of nodal lines] there will be a magnetic pole of strength [$g=n\hbar c/(2q)$].'' This is the original statement by which magnetic monopoles entered into the field of quantum mechanics.
 In 1948, Dirac \cite{3} presented a relativistic  extension of his theory of magnetic monopoles, in which he drew one of his most famous conclusions: ``Thus the mere existence of one pole of strength $[g]$ would require all electric charges to be quantised in units of $[\hbar c/ (2g)].$''

For the modern reader, the Dirac argument for the quantisation of the electric charge involving the elusive magnetic monopole is indeed ingenious. The basis of this argument is the interaction of an electric charge with the vector potential of a magnetic monopole attached to an infinitely long and infinitesimally thin solenoid, the so-called ``Dirac string" which is shown to be undetectable by assuming the single-valuedness of the wave function of the electric charge, and as a consequence the Dirac quantisation condition $q g=n\hbar c/2$ is required.  According to this condition, the existence of just one monopole anywhere in the Universe would explain why the electric charge is quantised. Indeed, if we identify the elementary magnetic charge with $g_0,$ then $q=n\hbar c/(2 g_0)$. Now for $n=1,$ we have the elementary electric charge $e=\hbar c/(2 g_0)$, which combines with $q=n\hbar c/(2 g_0)$ to give the law expressing the quantisation of the electric charge: $q=n e$. At the present time, the Dirac quantisation condition provides the most accepted explanation for the electric charge quantisation even when it relies on the existence of still undetected magnetic monopoles.
It is pertinent to note that there are excellent books \cite{4,5,6,7} and reviews \cite{8,9,10,11,12,13,14,15,16,17}
on magnetic monopoles, which necessarily touch on the subject of the Dirac quantisation condition and the Dirac string. So far, however, a review paper dealing with the Dirac condition rather than with magnetic monopoles seems not to appear in the standard literature.  The present review attempts to fill this gap for the benefit of the non-specialist.

Typically, the Dirac condition is discussed in graduate texts on electrodynamics \cite{18,19,20}, quantum mechanics \cite{21} and quantum field theory \cite{22,23,24,25}. The topic is rarely discussed in undergraduate textbooks \cite{26}. The purpose of this review is to survey the ideas underlying the Dirac quantisation condition, in a way that may be accessible to advanced undergraduate as well as graduate students. After commenting on the status of the Dirac quantisation condition, i.e., to discuss its past and present impact on theoretical physics, we find convenient to review the derivation of the Dirac condition given in Jackson's book \cite{18}. We next present a heuristic derivation of the this condition in which we attempt to follow Feynman's teaching philosophy that if we cannot provide an explanation for a topic at the undergraduate level then it means we do not really understand this topic \cite{27}. We then review four quantum-mechanical and three semi-classical derivations of the Dirac quantisation condition. Some of the relevant calculations involved in these derivations are detailed in Appendices. We think worthwhile to gather together the basic ideas underlying these derivations in a review, which may be accessible to advanced undergraduate and graduate students.

\section{Status of the Dirac quantisation condition: past and present}
\label{2}

\noindent To appreciate the relevance of the method followed by Dirac to introduce his quantization condition, let us briefly outline the historical context in which Dirac derived this condition. As is well known, Maxwell built his equations on the assumption that no free magnetic charges exist, which is formally expressed by the equation $\gradv\cdot\v B=0$. With the advent of quantum mechanics, magnetic charges were virtually excluded because the coupling of quantum mechanics with electrodynamics required the inclusion of the vector potential $\v A $ defined through $\v B=\gradv\times \v A.$ But it was clear that this equation precluded magnetic monopoles because of the well-known identity $\gradv\cdot(\gradv\times \v A)\!\equiv\!0.$ Before 1931, magnetic monopoles were irreconcilable within an electrodynamics involving the potential $\v A,$ and hence with quantum mechanics \cite{28}. Furthermore, for quantum physicists of the early twentieth century,  magnetic monopoles were mere speculations lacking physical content and were therefore not of interest at all in quantum theory prior to 1931. This was the state of affairs when Dirac suggested in his 1931 paper \cite{2} to reconsider the idea of magnetic monopoles. Using an innovative method, Dirac was able to reconcile the equations $\gradv\cdot\v B\not=0$ and $\v B=\gradv\times \v A,$ and therefore he was successful in showing that the interaction of an electron with a magnetic monopole was an idea fully consistent in both classical and quantum physics.

According to Dirac, the introduction of monopoles in quantum mechanics required magnetic charges to be necessarily quantised in terms of the electric charge and that quantisation of the latter should be in terms of the former. In his own words \cite{2}: ``Our theory thus allows isolated magnetic poles  [$g$], but the strength of such poles must be quantised, the quantum [$g_0$] being connected with the electronic charge $e$ by $[g_0=\hbar c/(2e)]$ ... The theory also requires a quantisation of electric charge ....'' In his 1931 paper \cite{2}, Dirac seems to favor the monopole concept when he pointed out: ``... one would be surprised if Nature had made no use of it.
''. As Polchinski has noted \cite{29}: ``From the highly precise electric charge
quantisation that is seen in nature, it is then tempting to infer that magnetic
monopoles exist, and indeed Dirac did so''. However, Dirac was very aware that isolated magnetic monopoles were still undetected and he proposed a physical explanation for this fact. When interpreting his result $g_0=(137/2)e$, he pointed out: ``This means that the attractive force between two one-quantum poles of opposite sign is $46921/4$ times that between electron and proton. This very large
force may perhaps account for why poles of opposite sign have never yet been separated.''

Let us emphasise that the true motivation of Dirac in his 1931 paper was twofold; on one hand, he wanted to explain the electric charge quantisation and on the other, to find the reason why the elementary electric charge had its reported experimental value. Such motivations were explicitly clarified by Dirac in 1978 \cite{30}: ``I was not searching for anything like monopoles at the time. What I was concerned with was the fact that electric charge is always observed in integral multiples of the electronic charge $e$, and I wanted some explanation for it. There must be some fundamental
reason in nature why that should be so, and also there must be some
reason why the charge $e$ should have just the value that it does have. It has the
value that makes $[\hbar c/e^2]$  approximately 137. And I was looking for some
explanation of this 137.''

In his 1948 paper \cite{3}, Dirac stressed the idea that each magnetic monopole is attached at the end of an ``unobservable'' semi-infinite string (a refinement of the nodal lines introduced in his 1931 paper \cite{2}). In retrospective, one can imagine that the idea of an unobservable string might have seemed strange at that time, and if additionally the theory was based on the existence of undetected magnetic monopoles, then it is not difficult to understand why this theory was received sceptically by some of Dirac's contemporaries. In a first view, Pauli disliked the idea of magnetic monopoles and
sarcastically referred to Dirac as ``Monopoleon''. But some years later, he reconsidered his opinion by saying that \cite{31}: ``This title [Monopoleon] shall indicate that I have a friendlier view to his theory of `monopoles' than earlier: There is some mathematical beauty
in this theory.'' On the other hand, Bohr, unlike Dirac, thought that one would be surprised if Nature had made use of magnetic monopoles \cite{32}.

After Dirac's 1931 seminal paper, Saha \cite{33} presented in 1936 a semi-classical derivation of the Dirac quantisation condition
based on the quantisation of the electromagnetic angular momentum associated to the static configuration formed by an electric charge and a magnetic charge separated by a finite distance, the so-called Thomson dipole (\cite{34}, see also \cite{35}). This same derivation was independently presented in 1949 by Wilson \cite{36,37}. In 1944, Fierz \cite{38} derived the Dirac condition by quantising the electromagnetic angular momentum arising from the classical interaction of a moving charge in the field of a stationary magnetic monopole. Schwinger \cite{39} in 1969 used a similar approach to derive a duality-invariant form of the Dirac condition by assuming the existence of particles possessing both electric and magnetic charges, the so-called dyons.

On the other hand, the Aharanov--Bohm effect \cite{40} suggested in 1959 has been recurrently used to show the undetectability of the Dirac string \cite{1,8,9,10,11,12,14,15,16,17,22,23,41,42}, giving a reversible argument. If Dirac's condition holds then the string is undetectable, and vice versa, if the string is undetectable then Dirac's condition holds. The path-integral approach to quantum mechanics, suggested by Dirac in 1933 \cite{43}, formally started by Feynman in his 1942 Ph.D. thesis \cite{44} and completed by him in 1948 \cite{48}, has also been used to obtain the Dirac condition \cite{22}.

Several authors have criticised the Dirac argument because of its unpleasant feature that it necessarily involves singular gauge transformations \cite{9}.
A formal approach presented by Wu and Yang \cite{46} in 1975 avoids such annoying feature by considering non-singular potentials, using the single-valuedness of the wave function and then deriving the Dirac condition without using the Dirac string \cite{4,8,9,11,12,13,16,24}. Other derivations of the Dirac condition have been presented over the years, including one by Goldhaber
\cite{47}, Wilzcek \cite{48,49} and Jackiw \cite{50,51,52}.

Remarkably, in 1974 t'Hooft \cite{53} and Polyakov \cite{54} independently discovered monopole solutions for spontaneously broken non-Abelian gauge theories. This originated another way to understand why electric charge is quantised in grand unified theories, where monopoles are necessarely present. If the electromagnetic $U(1)$ gauge group is embedded into a non-Abelian gauge group, then charge quantisation is automatic, for considerations of group theory \cite{4,11}. It is not surprising then that charge quantisation is now considered as an argument in support of grand unified theories \cite{4,29,55}.  In the context of unified theories, Polchinski goes even further arguing that \cite{29}  ``In any theoretical framework that requires charge to be quantised,
there will exist magnetic monopoles.'' On the other hand, it has been noted that the integer $n$ in Dirac's condition can be identified as a winding number, which gives a topological interpretation of this condition \cite{4,11,56}. Finally, it is pertinent to mention the recent claim that the Dirac condition also holds in the Proca electrodynamics with non-zero photon mass \cite{57}, reflecting the general character of this quantisation condition.

The preceding comments allow us to put in context the review presented here on the basic ideas underpinning the Dirac quantisation condition, such as gauge invariance, singular vector potentials, single-valuedness of the wave function, undetectibility of the Dirac string and the quantisation of the electromagnetic angular momentum.

The present review is organised as follows. In Section~\ref{3}, we closely review Jackson's treatment of the Dirac quantisation condition. In Sections~\ref{3}-\ref{6}, we present a new derivation of the Dirac condition based on heuristic and formal arguments, which does not consider the Dirac string. The specific gauge function required in this heuristic derivation is discussed. In Section~\ref{7}, we examine in detail the Dirac strings by explicitly identifying their singular sources. In Section~\ref{8}, we study the classical interaction of the electric charge with the Dirac string and conclude that this string has a mathematical rather than a physical meaning. In Section~\ref{9}, we examine the quantum-mechanical interaction of the electric charge with the Dirac string and show that if the string is undetectable then the Dirac quantisation condition holds. We review in Section~\ref{10} the Aharanov--Bohm effect and show how it can be used to derive the Dirac condition. In Section~\ref{11}, we outline Feynman's path integral approach to quantum mechanics and show how it can be used to obtain the Dirac condition. In Section~\ref{12}, we briefly discuss the Wu--Yang approach
that allows us to derive the Dirac condition without the recourse of the Dirac string. In Section~\ref{13}, we review three known semi-classical derivations of the Dirac condition. The first one makes use of the Thomson dipole. The second one considers the interaction between a moving charge and the field of a stationary monopole, and the third one considers the interaction between a moving dyon and the field of a stationary dyon. In Section~\ref{14}, we make some final remarks on the Dirac quantisation condition. In Section~\ref{15}, we make a final comment on the concept of nodal lines and in Section~\ref{16}, we present our conclusions. In Appendices \ref{A}--\ref{E}, we perform some calculations involved in the derivations of the Dirac condition.

\section{Jackson's treatment of the Dirac quantisation condition}
\label{3}

\begin{figure}[h]
  \centering
  \includegraphics[width=228pt]{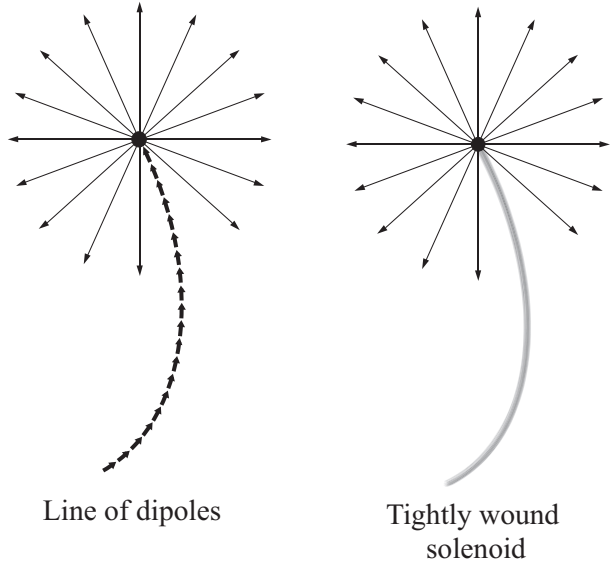}
  \caption{ Representation of a magnetic monopole $g$ as the end of a line of dipoles or as the end of a tightly wound solenoid that stretches off to infinity. }\label{Fig2}
\end{figure}

\noindent The {\it first quantum-mechanical derivation of the Dirac condition} we will review is that given in Jackson's book \cite{18}.
The magnetic monopole is imagined either as one particle to be at the end of a line of dipoles or at the end of a tightly wound solenoid that stretches off to infinity, as shown in Figure \ref{Fig2}. Any of these equivalent configurations can be described by the vector potential of a magnetic dipole $\v A(\v x)=[\v m\times (\v x- \v x')]/|\v x-\v x'|^3$,
where $\v x$ is the field point, $\v x'$ is the source point and $\v m$ is the magnetic dipole moment.
 The line of dipoles is a string formed by infinitesimal magnetic dipole moments $d\v m$  located at $\v x'$ whose vector potential is
$d\v A(\v x)=-d\v m\times\gradv\big(1/|\v x-\v x'|\big)$, where we have used $\gradv\big(1/|\v x-\v x'|\big)=-(\v x-\v x')/|\v x-\v x'|^3$. With the identification $d\v m= g d\v l'$, with $g$ being the magnetic charge and
$d\v l'$ a line element, the total vector potential for a string or solenoid lying on the curve $L$ reads
\begin{align}
\v A_L=-g\int_L d\v l'\times \gradv\bigg(\frac{1}{|\v x-\v x'|}  \bigg).
\end{align}
Using the result $\gradv \times (d \v l'/|\v x\!-\!\v x'|) = -d\v l' \times \gradv(1/|\v x\! -\!\v x'|)$, we can write Equation (1) as
\begin{align}
\v A_L=g\gradv\times \int_L \frac{d\v l'}{|\v x-\v x'|}.
\end{align}
Notice that this potential is already in the Coulomb gauge: $\gradv\cdot \v A_L=0$ because $\gradv\cdot [\gradv\times (\;\;)]\equiv 0$. 
\newpage
\noindent In Appendix \ref{A}, we show that the curl of this potential gives
\begin{align}
\gradv\times\v A_L=\frac{g}{R^2}\hat{\v R} + 4\pi g\!\int_L\! \delta(\v x-\v x')\,d\v l',
\end{align}
where $\delta(\v x-\v x')$ is the Dirac delta function, $R\!=\!|\v x-\v x'|$ and $\hat{\v R}\!=\!(\v x-\v x')/R$. To have a clearer meaning of Equation (3), it is convenient to write this equation as
\begin{align}
\v B_\texttt{mon}\!=\! \gradv\times \v A_L-\v B_\texttt{string},
\end{align}
where
\begin{align}
\v B_\texttt{mon}=\frac{g}{R^2}\hat{\v R},
\end{align}
is the field of the magnetic monopole $g$ located at the point $\v x'$ and
\begin{align}
\v B_\texttt{string}= 4\pi g\!\int_L \!\delta(\v x-\v x')\,d\v l',
\end{align}
is a singular magnetic field contribution along the curve $L$.

By taking the divergence to $\v B_\texttt{mon}$ it follows
\begin{align}
\gradv\cdot\v B_\texttt{mon}=&\gradv\cdot\bigg(\frac{g}{R^2}\hat{\v R}\bigg)= 4\pi g\delta(\v x\!-\!\v x'),
\end{align}
where we have used $\gradv \cdot (\hat{\v R}/R^2)=4 \pi \delta (\v x - \v x').$ Similarly, if we take the divergence to $\v B_\texttt{string},$ we obtain the result
\begin{align}
\nonumber \gradv\cdot\v B_\texttt{string}=& \gradv \cdot \bigg(4\pi g\!\int_L \!\delta(\v x-\v x')\,d\v l'\bigg)\\
=&\nonumber -4\pi g\!\int_L\! \gradv'\delta(\v x\!-\!\v x')\cdot d\v l'\\
=&-4\pi g\,\delta(\v x\!-\!\v x'),
\end{align}
where we have used $\gradv \delta(\v x\!-\!\v x')=-\gradv'\delta(\v x\!-\!\v x').$ When Equations (7) and (8) are used in the divergence of Equation (3) we verify the expected result $\gradv\cdot(\gradv\times\v A_L)=0$. Expressed in an equivalent way, the fluxes of the fields $\v B_\texttt{mon}$ and $\v B_\texttt{string}$ mutually cancel:
\begin{align}
\oint_S\v B_\texttt{mon}\cdot d\v a=&\int_V\gradv\cdot\v B_\texttt{mon}\, d^3x= 4\pi g,\\
\oint_S\v B_\texttt{string}\cdot d\v a=&\int_V\gradv\cdot\v B_\texttt{string}\,d^3x= -4\pi g,
\end{align}
where $d\v a$ and $d^3x$ denote the differential elements of surface and volume, and the Gauss theorem has been used. As a particular application, let
us consider the case in which the string lays along the negative $z$-axis and the magnetic monopole is at the origin. In this case $d\v l'=dz'\hat{\bf z}$, and the corresponding potential is
\begin{align}
\v A_L=g \gradv \times \hat{\v z}\int\limits_{-\infty}^{0}\frac{ dz'}{|\v x-z'\hat{\v z}|}.
\end{align}
In Appendix \ref{A}, we show that the curl of Equation (11) yields
\begin{align}
\gradv \times \v A_L =\frac{g}{r^2}\hat{\v r} +  4\pi g\delta(x)\delta(y)\Theta(-z)\hat{\v z},
\end{align}
where now $r=|\v x|, \hat{\v r}=\v x/r$, and $\Theta(z)$ is the step function which is undefined at $z=0$ but it is defined as $\Theta(z)\!=\!0$ if $z<\!0$ and $\Theta(z)\!=\!1$ if $z\!>\!0$. The highly singular character of the magnetic field of the string is clearly noted in the second term on the right of Equation (12). It is interesting to note that in his original paper \cite{2}, Dirac wrote the following solution for the vector potential in spherical coordinates $\v A_L= (g/r)\tan (\theta/2)\hat{\phi}$ and noted that this potential gives the radial field $g\hat{\v r}/r^2$. He pointed out: ``This solution is valid at all points except along the line $\theta = \pi$, where [$\v A_L$] become infinite.'' The solution considered by Dirac is
 equivalent to
\begin{align}
\v A_L= g\frac{1-\cos\theta}{r\sin\theta}\hat{\phi}.
\end{align}
This expression can be obtained by performing the integration specified in Equation (11), which requires the condition
$\sin\theta\not=0$. This is shown in Appendix \ref{B}.

Clearly, the curl of Equation (13) subjected to $\sin\theta\not=0$ gives only the field of the magnetic monopole  $\gradv\times \v A_L= g\hat{\v r}/r^2=\v B_\texttt{mon}$. This is so because the singularity originated by $\sin\theta=0$ is avoided in the differentiation process. As far as the computation of the total magnetic field of the configuration formed by a string laying along the negative $z$-axis and a magnetic monopole at the origin is concerned, it is simpler to take the curl to the implicit form of the potential defined by Equation (11) rather than taking the curl of a regularised form of the potential in Equation (13) [see Appendix \ref{D}].
\begin{figure}[h]
  \centering
  \includegraphics[width=230pt]{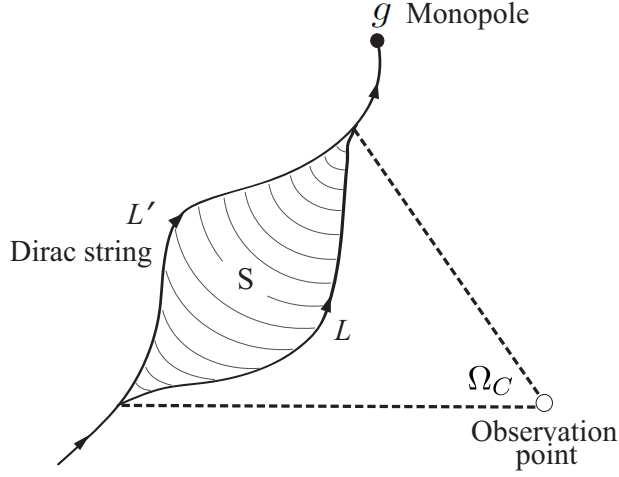}
  \caption{ Representation of a magnetic monopole $g$ as the end of a line of dipoles or as the end of a tightly wound solenoid that stretches off to infinity. The solid angle $\Omega_C$  is subtended by the curve $C = L - L'$,  which embeds the area $S.$
  The potentials $\v A_L$ and $\v A_{L'}$ correspond to the strings $L$ and $L'.$ }\label{Fig3}
\end{figure}
If an electric charge is interacting with the potential given in Equation (2), then it is ultimately interacting with a magnetic monopole and a magnetised string. Dirac argued that the interaction must only be with the magnetic monopole and therefore the charge $q$ should never ``see'' the singular field $\v B_\texttt{string}$ defined by Equation (6). For this reason he postulated  that the wave function must vanish along the string. But this requirement is certainly criticisable because it would mean that the string does not exist at all. This postulate is known as the ``Dirac veto" which in an alternative form states that any interaction of the electric charge with the string is forbidden. In Dirac's own words \cite{30}: ``You must have the monopoles and the electric charges occupying distinct regions of space. The strings, which come out from the monopoles, can be drawn anywhere subject to the condition that they must not pass through a region where there is electric charge present.''

The next step of the argument is to show that Equation (4) does not depend on the location of the string. To show this statement, consider two different strings $L'$ and $L$ with their respective vector potentials $\v A_{L'}$ and $\v A_L$.
 Evidently, the equivalence of these potentials will imply the equivalence of their respective strings indicating that the location of the string is irrelevant. The difference of the potentials $\v A_{L'}$ and $\v A_L$ can be obtained from Equation (2) with the integration taken along the closed curve $C=L'-L$ around the area $S$ as shown in Figure \ref{Fig3}. The result can be written as \cite{18}
\begin{align}
\v A_{L'}-\v A_L= g\gradv\times \oint_C \frac{d\v l'}{|\v x-\v x'|}= \gradv (g\Omega_C),
\end{align}
where $\Omega_C$ is the solid angle function subtended by the curve $C$. The integral specified in Equation (14) is done in Appendix \ref{C}. The fact that $\v A_{L'}$ and $\v A_{L}$ are connected by the gradient of a function reminds us of the gauge transformation $\v A'=\v A+ \gradv\Lambda$, where $\Lambda$ is a gauge function. Without any loss of generality, we can then write $\v A'\equiv\v A_{L'}, \v A\equiv\v A_{L}$ and $\Lambda\equiv g\Omega_C$. Notice that $\v A_{L'}$ and $\v A_{L}$ are in the Coulomb gauge: $\gradv \cdot\v A_{L'}=0$ and $\gradv \cdot\v A_{L}=0$. However, this does not prevent these potentials from being connected by a further gauge transformation whenever the gauge function $\Lambda$ is restricted to satisfy $\gradv^2\Lambda=0$. We can verify that this is indeed the case by taking the divergence to Equation (14) and obtaining $\gradv^2\Lambda=0,$ indicating that the potentials $\v A_{L'}$ and $\v A_{L}$ are connected by a restricted gauge transformation.

The remarkable point here is that different string positions correspond to different choices of gauge, or a change in string from $L$ to $L'$ is equivalent to a gauge transformation from $\v A_{L}$ to $\v A_{L'}$ with the gauge function $\Lambda=g\Omega_C$.  With the identification $\Lambda=g\Omega_C$, the associated phase transformation of the wave function $\Psi'= {\rm e}^{iq \Lambda/(\hbar c)}\Psi$ takes the form $\Psi'= {\rm e}^{iqg\Omega_C /(\hbar c)}\Psi.$ Now a crucial point of the argument. The solid angle $\Omega_C$ undergoes a discontinuous variation of $4\pi$ as the observation point (or equivalently the charge $q$) crosses the surface $S$. This makes the gauge function $\Lambda=g\Omega_C$ multi-valued which implies that ${\rm e}^{iqg\Omega_C}$ is also multi-valued, i.e.,  ${\rm e}^{iqg\Omega_C}\!\not=\!{\rm e}^{iqg(\Omega_C+4\pi)}.$ Thus the transformed wave function of the charge $q$ will be multi-valued when $q$ crosses $S$, unless we impose the condition  ${\rm e}^{i4\pi qg /(\hbar c)}\!=\!1\!$.  But this condition and
${\rm e}^{i 2\pi n}=1$  with $n$ being an integer, imply  $4\pi qg/(\hbar c)\! =\!2\pi n$, and hence, the Dirac quantisation condition $q g=n\hbar c/2$ is obtained. Accordingly, the field of the monopole in Equation (4) does not depend on the location of the string. The  price we must pay is the imposition of the Dirac condition. The lesson to be learned here is that gauge invariance and single-valuedness of the wave function are the basic pieces to ensemble the Dirac quantisation condition.

The above derivation of the Dirac condition puts emphasis on the idea that the location of the string is irrelevant. But the argument might equally put emphasis on the idea that the string is unobservable. In fact, consider the value $\Omega_1$ corresponding to one side of the surface $S$ and the value
$\Omega_2$ corresponding to the other side. They are related by $\Omega_1=\Omega_2+4\pi$. It follows that ${\rm e}^{iqg\Omega_1/(\hbar c)}={\rm e}^{iqg(\Omega_2 +4\pi)/(\hbar c)}.$ This means that the wave function of the charge $q$ differs by the quantity ${\rm e}^{i4\pi qg/(\hbar c)}$, and this would make the Dirac string observable as the charge crosses the surface, unless we impose the condition ${\rm e}^{i4\pi qg/(\hbar c)}=1$, which is satisfied if $q g=n\hbar c/2$ holds, i.e, the price we must pay for the unobservability of the string is the imposition of the Dirac condition.

The standard derivation of the Dirac quantisation condition explained in this section is appropriate to be presented to graduate students. In  Sections \ref{4}-\ref{9} we will suggest a presentation of the Dirac condition that encapsules the main ideas underlying this condition, which may be suitable for advanced undergraduate students.

\section{How to construct a suitable quantisation condition}
\label{4}

\noindent
The origin of the letter $n$ appearing in the Dirac quantisation condition  $qg=n\hbar c/2$ can be traced to the trigonometric identity $\cos{(2\pi n)}=1,$ where $n=0\pm1,\,\pm2,\,\pm3 ...$ This trigonometric identity can be expressed as
\begin{align}
{\rm e}^{i2\pi n}=1,
\end{align}
which follows from Euler's formula ${\rm e}^{i\alpha}=\cos\alpha +i\sin\alpha$ with $\alpha=2\pi n$. Consider now spherical coordinates $(r,\theta,\phi)$ with their corresponding unit vectors $(\hat {\bf r},\hat{\theta},\hat{\phi})$. For fixed $r$ and $\theta$, the azimuthal angles $\phi$ and $\phi +2\pi$ represent the same point. This property allows us to define a single-valued function of the azimuthal angle $F(\phi)$ as one that satisfies $F(\phi)=F(\phi +2\pi)$.
We note that the particular function $F(\phi)=\phi$ is not a single-valued function because $F(\phi)=\phi$ and
$F (\phi +2\pi)=\phi+2\pi$ take different values: $F(\phi)\not=F(\phi+2\pi)$. We then say that $F=\phi$ is a multi-valued function.

The complex function $F(\phi)={\rm e}^{i2k\phi}$ with $k$ being an arbitrary constant is not generally a single-valued function because $F(\phi)={\rm e}^{i2k\phi}$ and   $F (\phi+2\pi)={\rm e}^{i2k(\phi +2\pi)}$ can take different values: $F(\phi)\not=F(\phi +2\pi)$. This is so because in general ${\rm e}^{i4\pi k}\not=1$ for arbitrary $k$. In this case, however, we can impose a condition on the arbitrary constant $k$ so that $F={\rm e}^{i2k\phi}$ becomes a single-valued function. By considering Equation (15), we can see that ${\rm e}^{i4\pi k}=1$ holds when $k$ is dimensionless and satisfies the ``quantisation'' condition:
\begin{align}
k=\frac{n}{2}, \quad n=0,\pm1,\pm2,\pm3,....
\end{align}
Under this condition, $F={\rm e}^{i2k\phi}$ becomes a single-valued function: $F(\phi)=F(\phi +2\pi)$. In short: the single-valuedness of $F={\rm e}^{i2k\phi}$ requires the quantisation condition specified in Equation (16). Notice that a specific value of $k$  may be obtained in principle by considering the basic equations of a specific physical theory. We will see that electrodynamics with magnetic monopoles and quantum mechanics conspire to yield the specific value of $k$ that leads to the Dirac quantisation condition.

\section{Gauge invariance and the Dirac quantisation condition}
\label{5}
\noindent
We will now to present a {\it heuristic quantum-mechanical derivation of the Dirac condition}.
The Schr\"odinger equation for a non-relativistic particle of mass $m$ and electric charge $q$ coupled to a time-independent vector potential $\v A(\v x)$ is given by
\begin{align}
i \hbar \frac{\partial \Psi}{\partial t}= \frac{1}{2m}\bigg(\!-i\hbar \gradv - \frac{q}{c}\v A \bigg)^{\!2}\Psi.
\end{align}
This equation is invariant under the simultaneous application of the gauge transformation of the potential
\begin{align}
\v A' = \v A+ \gradv \Lambda,
\end{align}
and the local phase transformation of the wave function
\begin{align}
\Psi'= {\rm e}^{iq \Lambda/(\hbar c)}\,\Psi,
\end{align}
where $\Lambda(\v x)$ is a time-independent gauge function. Equations (17)-(19) are well known in textbooks (see note at the end of this review).

At first glance, Equations (17)-(19) do not seem to be related to some quantisation condition. But a comparison between the previously discussed function ${\rm e}^{i 2k\phi}$ with the phase factor
${\rm e}^{iq\Lambda/(\hbar c)}$ appearing in Equation (19),
\begin{align}
{\rm e}^{i 2k\phi}\;\;\longleftrightarrow \;\;{\rm e}^{iq\Lambda/(\hbar c)},
\end{align}
suggests the possibility of constructing a specific quantisation condition connected with Equations (17)-(19). Consider first that $k$ is an arbitrary constant. Therefore ${\rm e}^{i 2 k\phi}$ is not generally a single-valued function. We recall that the gauge function $\Lambda$ in the phase ${\rm e}^{iq\Lambda/(\hbar c)}$ of the transformation in Equation (19) is an arbitrary function which may be single-valued or multi-valued. In view of the arbitrariness of $k$ and
$\Lambda$, we can make equal both functions: ${\rm e}^{iq \Lambda/(\hbar c)}={\rm e}^{i2 k\phi}$, which implies
\begin{align}
\Lambda q=2 k \hbar c \phi.
\end{align}
This is the key equation to find a quantisation condition that leads to the electric charge quantisation. The genesis of this remarkable equation is
the gauge invariance of the interaction between the charge $q$ and the potential $\v A$. By direct substitution we can show that a particular solution of Equation (21) is given by the relations
\begin{align}
k=\frac{qg}{\hbar c},
\end{align}
and
\begin{align}
\Lambda= 2g \phi,
\end{align}
where the constant $g$ is introduced here to make the constant $k$ dimensionless. The constant $g$ has the dimension of electric charge and its physical meaning is unknown at this stage. Notice that $\Lambda$ in Equation (23) is a multi-valued gauge function. We require now that the phase ${\rm e}^{iq \Lambda/(\hbar c)}$ be single-valued. From ${\rm e}^{iq \Lambda/(\hbar c)}={\rm e}^{i 2k\phi}$ it follows that ${\rm e}^{i 2k\phi}$ must be single-valued and then $k$ must satisfy the quantisation condition displayed in Equation (16). In other words, by demanding the single-valuedness of
${\rm e}^{iq \Lambda/(\hbar c)}$, Equations (16) and (22) yield the quantisation condition
\begin{align}
qg=\frac{n}{2}\hbar c.
\end{align}
If now the constant $g$ is assumed to be the magnetic charge then Equation (24) is the Dirac quantisation condition.

Notice that according to the heuristic approach followed here, the derivation of Equation (24) relies on the existence of the gauge function $\Lambda= 2g \phi$.  In the following section we will discuss the feasibility of this specific gauge function and argue the identification of $g$ with the magnetic charge. For now we observe that the heuristic approach uses the same two fundamental pieces discussed in Section~\ref{3}, namely, the single-valuedness of the wave function and gauge invariance. However,  the heuristic approach makes use of these two pieces in a simpler way.

\section{The gauge function $\Lambda= 2g \phi$}
\label{6}
\noindent It is convenient to assume first the existence of the gauge function $\Lambda= 2g \phi$ with the purpose of elucidating its associated gauge potentials. The gradient  of $\Lambda= 2g \phi$ in spherical coordinates gives
\begin{align}
\gradv \Lambda=\frac{2g}{r\sin{\theta}}\hat{\phi}.
\end{align}
Notice that this gradient is singular at $r=0$. This is a real singular point which is not problematic and we agree it is allowed. However, this gradient is also singular at those values of the polar coordinate $\theta$ satisfying $\sin{\theta}=0$,
which represent lines of singularities involving non-trivial consequences, which will be discussed in Section~\ref{7}. Presumably, there exist two vector potentials such that
\begin{align}
\v A'-\v A=\frac{2g}{r\sin{\theta}}\hat{\phi}.
\end{align}
Both potentials $\v A'$ and $\v A$ must originate the same magnetic field $\v B$, i.e.,  $\gradv\times\v A'=\gradv\times\v A=\v B$. From Equation (26) we can see that
 $\v A'$ and $\v A$ may be of the generic form
\begin{align}
\v A'= g\frac{1-f(\theta)}{r\sin\theta}\hat{\phi}, \quad \v A= -g\frac{1+f(\theta)}{r\sin\theta}\hat{\phi},
\end{align}
where $f(\theta)$ is an unspecified function such that it does not change the validity of Equation (26). Notice that $\v A'$ and $\v A$ have singularities originated by $\sin \theta\!=\!0$. These will not be considered for now. We observe that $\v A'$ and $\v A$ in Equation (27) are of the form $\v A'=[0,0,A'_\phi(r,\theta)]=A'_\phi(r,\theta)\hat{\phi}$ and  $\v A=[0,0,A_\phi(r,\theta)]=A_\phi(r,\theta)\hat{\phi}$. The curl of a generic vector of the form $\bfF=\bfF[0,0,F_\phi(r,\theta)]$ in spherical coordinates reads
\begin{align}
\gradv\times \bfF=\frac{1}{r\sin{\theta}}\frac{\partial}{\partial \theta}\big(\sin{\theta}F_\phi\big)\hat{\v r}-\frac{1}{r}\frac{\partial}{\partial r}\big(r F_\phi\big)\hat{\theta}.
\end{align}
When this definition is applied to $\v A'$ and $\v A$ and $\sin \theta\not=0$ is assumed we obtain
\begin{align}
\gradv\times \v A'=\gradv\times \v A=-\frac{g}{r^2\sin{\theta}}\frac{\partial f}{\partial \theta}\hat{\v r},
\end{align}
and therefore both potentials yield the same field
\begin{align}
\v B=-\frac{g}{r^2\sin{\theta}}\frac{\partial f}{\partial \theta}\hat{\v r}.
\end{align}
In the particular case $f(\theta)=\cos\theta,$ this field becomes
\begin{align}
\v B=\frac{g}{r^2}\hat{\v r}.
\end{align}
 The nature of the constant $g$ is then revealed in this particular case. Equation (31) is the magnetic field produced by a magnetic charge $g$ located at the origin. In other words, the constant $g$ introduced by hand in Equations (22) and (23) is naturally identified with the magnetic monopole!

The potentials $\v A'$ and $\v A$ in Equation (27) are in the Coulomb gauge. In fact, using the definition of the divergence of the generic vector $\bfF=\bfF[0,0,F_\phi(r,\theta)]$ in spherical coordinates $\gradv\cdot \bfF=[1/(r\sin\theta)]\partial \bfF_\phi/\partial \phi$, it follows that $\gradv\cdot \v A'=0$ and $\gradv\cdot \v A=0$. Here, there is a point that requires to be clarified. At first glance, there seems to be some inconsistency when connecting $\v A'$ and $\v A$ via a gauge transformation because both potentials are already in a specific gauge, namely, the Coulomb gauge. However, there is no inconsistence  as explained in Section \ref{3}, because even for potentials satisfying the Coulomb gauge there is arbitrariness. Evidently, the restricted gauge transformation $\v A\to \v A'=\v A+\gradv\Lambda$, where $\gradv^2\Lambda=0$, preserves the Coulomb gauge. The definition of the Laplacian of the generic scalar function $f=f(\phi)$ in spherical coordinates reads $\gradv^2 f=[1/(r\sin\theta)^2]\partial^2f/\partial \phi^2$. Using this definition with $f=\Lambda= 2g \phi,$ it follows that $\gradv^2\Lambda=0,$ indicating that  $\v A'$ and $\v A$ are connected by a restricted gauge transformation.

Let us recapitulate. By assuming the existence of the gauge function $\Lambda=2 g\phi$, we have inferred the potentials
\begin{align}
\v A'= g\frac{1-\cos\theta}{r\sin\theta}\hat{\phi}, \quad \v A= -g\frac{1+\cos\theta}{r\sin\theta}\hat{\phi}.
\end{align}
[these are $\v A'$ and $\v A$ in Equation (27) with $f(\theta)=\cos\theta$], which originate the same field given in Equation (31) whenever the condition $\sin \theta\not=0$ is assumed. This field is the Coloumbian field due to a magnetic monopole $g$.  With the identification of $g$ as the magnetic monopole, we can say that Equation (24) is the Dirac quantisation condition. Evidently, we can reverse the argument by introducing first the potentials $\v A'$ and $\v A$ by means of Equation (32) considering $\sin \theta\not=0$
and then proving they yield the same magnetic field in Equation (31). The existence of these potentials guarantees the existence of the gauge function $\Lambda=2 g\phi$.

Once the existence of the gauge function $\Lambda=2 g\phi$ has been justified with $g$ being the magnetic monopole, the heuristic derivation of the Dirac quantisation condition has been completed. However, we should note that this heuristic procedure involves an aspect that could be interpreted as an inconsistency. According to the traditional interpretation, the existence of magnetic monopoles implies $\gradv\cdot\v B\not=0$ and therefore we cannot write $\v B=\gradv\times \v A$, at least not globally. This is so because $\gradv\cdot(\gradv\times\v A)=0$. The origin of this apparent inconsistency deals with the singularity originated by the value $\sin \theta=0$ and its explanation will take us to one of the most interesting concepts in theoretical physics, the Dirac string, which will be discussed in the following section.

\section{Dirac strings}
\label{7}

\noindent As previously pointed out, both potentials in Equation (32) yield the same magnetic field given in Equation (31) whenever $\sin \theta\not =0$ is assumed. The question naturally arises: What does $\sin \theta =0$ mean? The answer is simple: $\theta=0$ and $\theta=\pi$. The first value represents the positive semi-axis $z$, i.e., $z>0$, whereas the second value represents the negative semi-axis $z$, i.e., $z<0$. Therefore, the condition $\sin \theta\not=0$ means that the semi-axes $z>0$ and $z<0$ have been excluded in the heuristic treatment. Accordingly, when we took the curl to $\v A'$ and $\v A$, we obtained the magnetic field $\v B=g\hat{\v r}/r^2$ in all space except at $r=0$ (which we agree it is allowed) and except along the negative semi-axis in the case of $\v A'$, and also except along the positive semi-axis in the case of $\v A$. Expectably, if we additionally consider the field contributions associated to the Dirac strings located in the positive and negative semi-axes then we can reasonably assume the following equations:
\begin{align}
\gradv \times\v A'= &\;\frac{g}{r^2}\hat{\v r} + \v B'({\rm along}\;z<0),\\
\gradv \times\v A = &\;\frac{g}{r^2}\hat{\v r} + \v B({\rm along}\;z>0).
\end{align}
Here $\v B'(z<0)$ and  $\v B(z>0)$ represent magnetostatic fields produced by Dirac strings. The formal determination of these
fields is not an easy task because they are highly singular objects. But, fortunately, heuristic considerations allow us to elucidate the explicit form of these fields. We note that the semi-axis $z<0$
can be represented by the singular function $-\delta(x)\delta(y)\Theta(-z)\hat{\v z}$ and the semi-axis $z>0$ by the singular function $\delta(x)\delta(y)\Theta(z)\hat{\v z}$. Therefore, the fields $\v B'(z<0)$ and $\v B(z>0)$ may be appropriately modelled by the singular functions
\begin{align}
\v B'(z<0)=& -K\delta(x)\delta(y)\Theta(-z)\hat{\v z},\\
\v B(z>0)=&\; K\delta(x)\delta(y)\Theta(z)\hat{\v z},
\end{align}
where $K$ is a constant to be determined. Using Equations (33)-(36), we obtain
\begin{align}
\gradv \times\v A'= &\;\frac{g}{r^2}\hat{\v r} -  K\delta(x)\delta(y)\Theta(-z)\hat{\v z},\\
\gradv \times\v A = &\;\frac{g}{r^2}\hat{\v r} + K\delta(x)\delta(y)\Theta(z)\hat{\v z}.
\end{align}
 The divergence of Equation (37) gives
 \begin{align}
0= 4\pi g\delta(\v x)+K\delta(\v x),
\end{align}
where $\gradv\cdot (\hat{\v r}/r^2)\!=\!  4\pi\delta(\v x)$ with $\delta(\v x)=\delta(x)\delta(y)\delta(z)$ and $\partial\Theta(-z)/\partial z\!=\!-\delta(z)$ have been used.
 A similar calculation on Equation (38) gives Equation (39) again. From Equation (39), it follows that $K=-4\pi g$ and thus we get the final expressions
 \begin{align}
\gradv \times\v A'= &\;\frac{g}{r^2}\hat{\v r} +  4\pi g\delta(x)\delta(y)\Theta(-z)\hat{\v z},\\
\gradv \times\v A = &\;\frac{g}{r^2}\hat{\v r} - 4\pi g\delta(x)\delta(y)\Theta(z)\hat{\v z}.
\end{align}
We should emphasise that simple heuristic arguments have been used to infer Equations (40) and (41). We also note that Equation (40) is the same as Equation (12), which was in turn derived by the more complicated approach outlined in Section~\ref{3}. The advantage of the heuristic argument is that it has nothing to do with the idea of modelling a magnetic monopole either as the end of an infinite line of infinitesimal magnetic dipoles or as the end of a tightly wound solenoid that stretches off to infinity. Equation~(40) is also formally derived in Appendix \ref{A} by means of an integration process. Furthermore, Equation (40) can alternatively be obtained by differentiation, which is done in Appendix \ref{D}, where an appropriate regularisation of the potential $\v A'$ is required.

\begin{figure}
  \centering
  \includegraphics[width= 268pt]{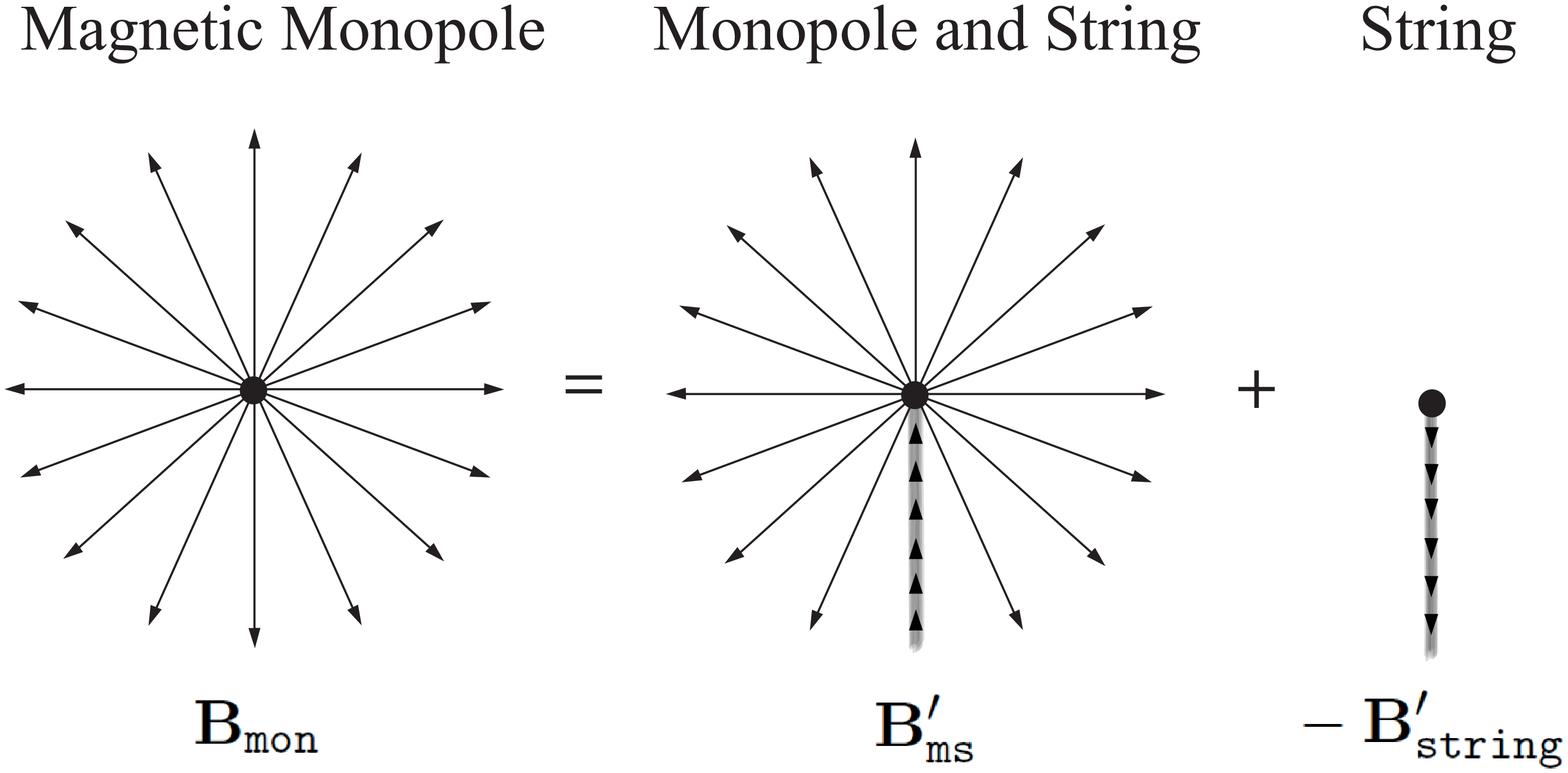}
  \caption{Pictorial representation of the monopole field $\v B_\texttt{mon}$ defined by Equation (42). We have extracted the field of the string $\v B'_\texttt{string}$ from the field $\v B'_\texttt{ms}$ to insolate the field $\v B_\texttt{mon}$ of the magnetic monopole.}\label{Fig4}
\end{figure}

\begin{figure}
  \centering
  \includegraphics[width= 268pt]{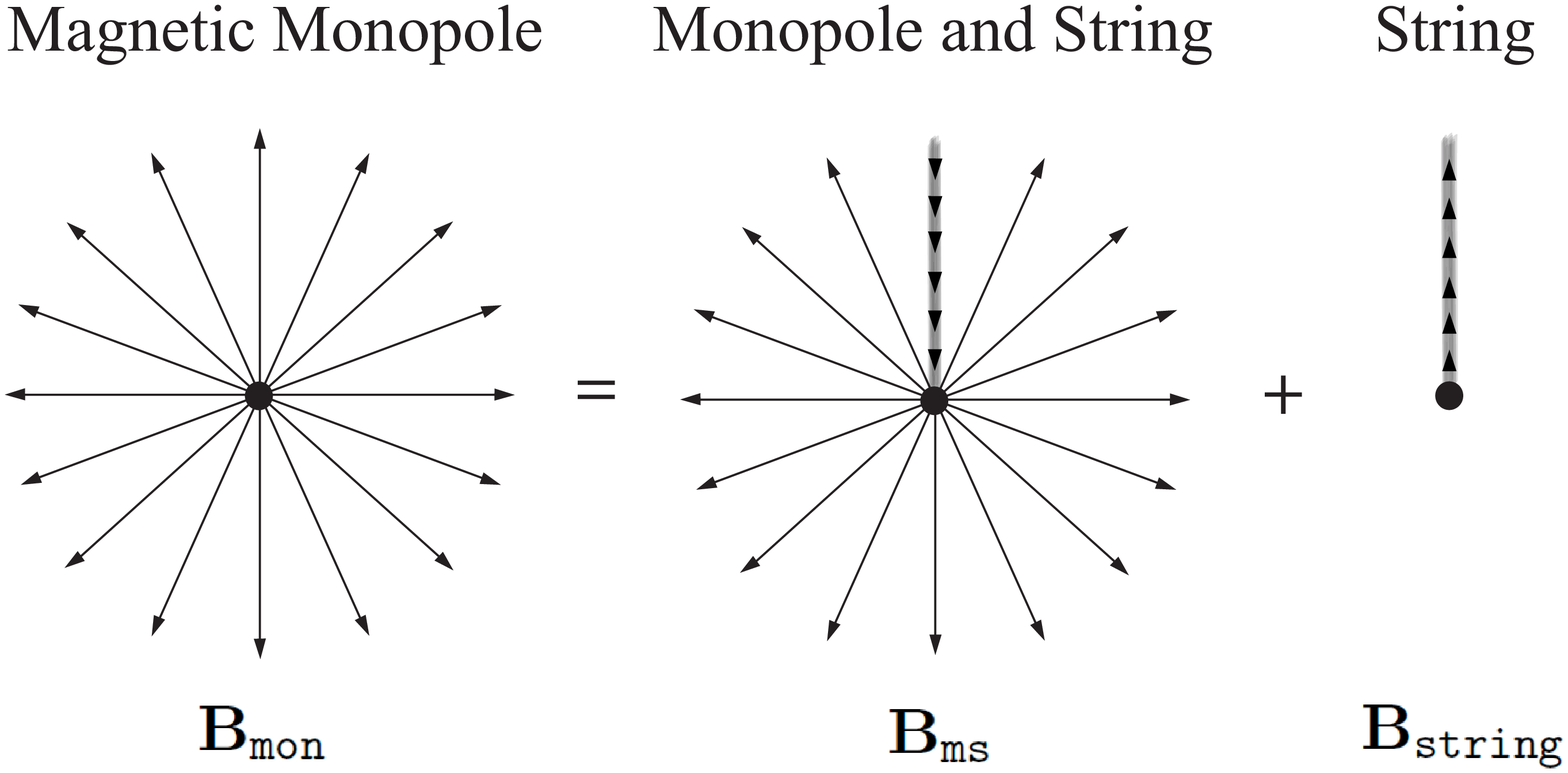}
  \caption{Pictorial representation of the monopole field $\v B_\texttt{mon}$ defined by Equation (43). We have added the field of the string $\v B_\texttt{string}$ to the field $\v B_\texttt{ms}$ to insolate the field $\v B_\texttt{mon}$ of the magnetic monopole.}\label{Fig5}
\end{figure}

Expressed differently, the potentials $\v A'$ and $\v A$ appearing in Equations (40) and (41) produce respectively the fields $\v B'_\texttt{ms}=\gradv \times\v A'$ and $\v B_\texttt{ms}=\gradv \times\v A$, and so we can write
\begin{align}
\v B'_\texttt{ms}= &\;\v B_\texttt{mon} +\v B'_\texttt{string},\\
\v B_\texttt{ms}= &\;\v B_\texttt{mon} +\v B_\texttt{string},
\end{align}
where the respective magnetic fields are defined as
\begin{align}
\v B_\texttt{mon}=&\;\frac{g}{r^2}\hat{\v r},\\
\v B'_\texttt{string}=&\;4\pi g\delta(x)\delta(y)\Theta(-z)\hat{\v z},\\
\v B_\texttt{string}=&-  4\pi g\delta(x)\delta(y)\Theta(z)\hat{\v z}.
\end{align}
Figures \ref{Fig4} and \ref{Fig5} show a pictorial representation of the fields appearing in Equations (42) and (43).
It is conceptually important to identify the sources of the fields described by Equations (42) and (43). The magnetic field $\v B_\texttt{mon}$ in Equation (44)
satisfies
\begin{align}
\gradv\cdot\v B_\texttt{mon}&=4\pi g\delta(\v x),\\
\gradv\times\v B_\texttt{mon}&=0,
\end{align}
The magnetic field $\v B'_\texttt{string}$ in Equation (45) satisfies
\begin{align}
\gradv\cdot\v B'_\texttt{string}=& -4\pi g\delta(\v x), \\
\gradv\times\v B'_\texttt{string}=&\,4\pi g\Theta(-z)\big[\delta(x)\delta'(y) \hat{\v x}-\delta'(x)\delta(y)\hat{\v y}\big],
\end{align}
where $\delta'(x)= d\delta(x)/dx$ and $\delta'(y)= d\delta(y)/dy$ are delta function derivatives. The field $\v B_\texttt{string}$ in Equation (46) is shown to satisfy
\begin{align}
\gradv\cdot\v B_\texttt{string}&=-4\pi g\delta(\v x),\\
\gradv\times\v B_\texttt{string}&=-4\pi g\Theta(z)\big[\delta(x)\delta'(y) \hat{\v x}-\delta'(x)\delta(y)\hat{\v y}\big].
\end{align}
Therefore, the field  $\v B'_\texttt{ms}$  defined by Equation (42) satisfies
\begin{align}
\gradv\cdot \v B'_\texttt{ms}&= 0,
\end{align}
\begin{align}
\gradv\times\v B'_\texttt{ms}&= 4\pi g\Theta(-z)\big[\delta(x)\delta'(y) \hat{\v x}-\delta'(x)\delta(y)\hat{\v y}\big],
\end{align}
and the field  $\v B_\texttt{ms}$  defined by Equation (43) satisfies
\begin{align}
\gradv\cdot \v B_\texttt{ms}&= 0,\\
\gradv\times\v B_\texttt{ms}&= -4\pi g\Theta(z)\big[\delta(x)\delta'(y) \hat{\v x}-\delta'(x)\delta(y)\hat{\v y}\big].
\end{align}

Let us return to the Schr$\ddot{\rm o}$dinger equation defined by Equation (17). According to this equation, the electric charge $q$ interacts with the  potential $\v A$. From the gauge function  $\Lambda= 2g \phi,$ we inferred the potentials $\v A'$ and $\v A$ given in Equation (32). The curl of each of these potentials originates the field of the magnetic monopole plus the field of the respective string as may be seen in Equations (40) and (41). If any of these potentials is considered in Equation (17), then a question naturally arises: Does the electric charge interact only with the monopole or with the monopole and a Dirac string? In other words: Can the electric charge physically interact with a Dirac string? The answer is not as simple as might appear
at first sight.  The Dirac string is a subtle object whose physical nature has originated controversy and debate.

Typically, the magnetic field of the Dirac string is discussed together with the Coulombian field of the magnetic monopole. But since we have identified the sources of the magnetic field of the string [those given on the right of Equations (49) and (50) or also on the right of Equations (51) and (52)], we can study the magnetic field of the Dirac string with no reference to the Coulombian field. In the following section, we will discuss the interaction of an electric charge with a Dirac string from classical and quantum-mechanical viewpoints.

\section{Classical interaction between the electric charge and the Dirac string}
\label{8}
\noindent In order to understand the possible meaning of the Dirac string, we should first study the sources of the magnetostatic field produced by this string.  Let us assume that the string lies along the negative $z$-axis. From Equations (49) and (50), we can see that this string has the associated charge and current densities:
\begin{align}
\rho_\texttt{string}&=-g\delta(\v x),\\
\v J_\texttt{string}&=cg\Theta(-z)\big[\delta(x)\delta'(y) \hat{\v x}-\delta'(x)\delta(y)\hat{\v y}\big],
\end{align}
which generate the magnetic field
\begin{align}
\v B'_\texttt{string}= 4\pi g\delta(x)\delta(y)\Theta(-z)\hat{\v z}.
\end{align}
A regularised vector potential in cylindrical coordinates for the field $\v B'_\texttt{string}$ reads
 \begin{align}
\v A_\texttt{string}=\frac{2g\Theta(\rho-\varepsilon)\Theta(-z)}{\rho}\hat{\phi},
\end{align}
where $\varepsilon >0$ is an infinitesimal quantity.  Notice that the potential $\v A_\texttt{string}$ for $\rho>\varepsilon$ and $z<0$ is a pure gauge potential, i.e., it can be expressed as the gradient of a scalar field. To show that $\v A_\texttt{string}$ generates $\v B'_\texttt{string}$ consider the curl of the generic vector $\bfF= \bfF[0, F_\phi(\rho,z),0]$ in cylindrical coordinates
\begin{align}
\gradv\times \bfF=-\frac{\partial  F_\phi}{\partial z} \hat{\rho} + \frac{1}{\rho}\frac{\partial}{\partial \rho}\big(\rho F_\phi\big)\hat{\v z}.
\end{align}
When this definition is applied to the potential $\v A_\texttt{string}$ defined by Equation (60), we obtain
\begin{align}
\gradv\times\v A_\texttt{string}=&\, \frac{2g\Theta(\rho-\varepsilon)\delta(z)}{\rho}\hat{\rho} + \frac{2g\delta(\rho-\varepsilon)\Theta(-z)}{\rho}\hat{\v z}.
\end{align}
Since we are only considering $z\!<\!0$ the first term vanishes and then
\begin{align}
\gradv\times\! \v A_\texttt{string}&=\frac{2g\delta(\rho-\varepsilon)\Theta(-z)}{\rho}\hat{\v z}\nonumber\\
&= 4\pi g\delta(x)\delta(y)\Theta(-z)\hat{\v z}\nonumber\\
&=\v B'_\texttt{string},
\end{align}
where we have used the formula \cite{58}:
 \begin{align}
\delta(x)\delta(y)=\frac{\delta(\rho-\varepsilon)}{2\pi\rho},
\end{align}
in which the limit $\varepsilon\to 0$ is understood.

Having all the classical ingredients on the table, we will now proceed to interpret them from both mathematical and physical point of views.
These ingredients are highly singular and therefore such interpretations are full of subtleties. Assuming the existence of magnetic monopoles, the classical interaction between a moving electric charge $q$ and the magnetic field $\v B'_\texttt{string}$ is given by the Lorentz force $\v F = q(\bfv/c)\times \v B'_\texttt{string}$. Expressing the velocity $\bfv$ of the charge in cylindrical coordinates $\bfv=(v_\rho, v_\phi,v_z)$ and using the regularised form of $\v B'_\texttt{string}=\gradv \times \v A_\texttt{string}$ defined in the first line of Equation (63), this force reads
\begin{align}
\v F= -\frac{2qg\Theta(-z)}{c}\frac{\delta(\rho-\varepsilon)}{\rho}\big[v_\phi \hat{\rho}-v_\rho\hat{\phi}\big].
\end{align}
The singular character of this force becomes evident. If $\rho \neq\varepsilon $ this force vanishes and then the charge $q$ is insensitive to the string. If the charge $q$ approaches too much to the string, then $\rho\to \varepsilon$, which implies $\rho\to 0$ because $\varepsilon\to 0$. In this case, we have
 \begin{align}
\lim_{\rho\to 0} \frac{\delta(\rho-\varepsilon)}{\rho}=0,
\end{align}
and again the force in Equation (65) vanishes indicating that the charge $q$ is also unaffected by the string in this extreme case. However, from a mathematical point of view, when $\rho=\varepsilon$ the force in Equation (65) becomes infinite $(\infty/0=\infty),$ which is physically unacceptable.

Two results are then conclusive. On one hand, if the electric charge $q$ is outside the string, then $q$ does not feel the action of the magnetic field of the string. This is true even when the charge $q$ is very close to the string. On the other hand, if $\rho=\varepsilon,$ then the charge $q$ feels an infinite force due to the magnetic field of the string. The idea of an infinite force leads us to
conclude that the Dirac string lacks any physical meaning. Thus the common statement that the Dirac string cannot be detected is meaningful in purely classical considerations.

\begin{figure}
  \centering
  \includegraphics[width= 224pt]{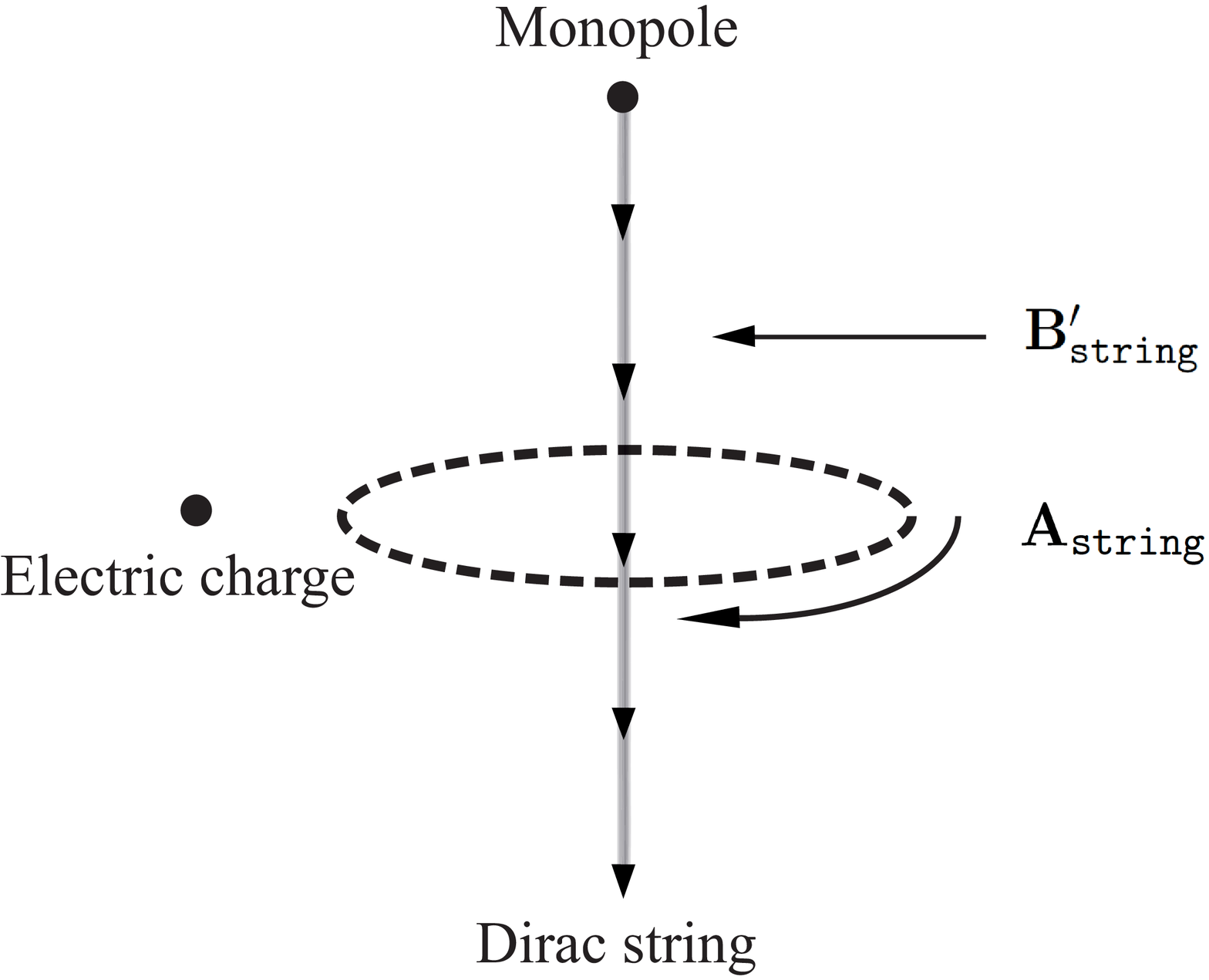}
  \caption{Geometry of the Dirac string and its associated  vector potential $\v A_\texttt{string}.$ This potential satisfies $\gradv \times \v A_\texttt{string}= \v B'_\texttt{string}.$}\label{Fig6}
\end{figure}

The interpretation of the potential in Equation (60) is also somewhat subtle. There is no problem when $\rho>\varepsilon$ because in this case
$\v A_\texttt{string}=2g\Theta(-z)\hat{\phi}/\rho$ exhibits a regular behaviour which is drawn in Figure \ref{Fig6}.
There is also no problem when  $\rho<\varepsilon$ because in this case $\v A_\texttt{string}=0$. When $\rho\to\varepsilon,$ it follows $\rho\to 0$ because $\varepsilon\to 0$. In this case
 \begin{align}
\lim_{\rho\to 0} \frac{\Theta(\rho-\varepsilon)}{\rho}=0,
\end{align}
and again $\v A_\texttt{string}$ vanishes.  The problematic issue arises when $\rho=\varepsilon$ because in this case $\v A_\texttt{string}$ becomes undefined.

\section{Quantum-mechanical interaction between the electric charge and the Dirac string}
\label{9}
\noindent The {\it second quantum-mechanical derivation of the Dirac condition} will now be reviewed.  We have argued that the classical interaction of an electric charge with the Dirac string is not physically admissible. Now we will consider the possibility of  a quantum-mechanical interaction between
the electric charge and the string. Dirac \cite {2} noted that the interaction of an electric charge with a vector potential is given by the phase in the wave function
 \begin{align}
\Psi={\rm e}^{i[q/(\hbar c)]\int_0^{\v x}\v A(\v x')\cdot\, d\v l'}\Psi_0,
\end{align}
where $\Psi_0$ is the solution of the free Schr\"odinger  equation and the line integral is taken a long a path of the electric charge from the origin to the point $\v x$. The quantum mechanical analogous to the classical Lorentz force $\v F = q(\bfv/c)\times \v B$ is given by the phase ${\rm e}^{i[q/(\hbar c)]\int_0^{\v x}\v A(\v x')\cdot d\v l'}$ appearing in Equation (68),
 which in turn represents the solution of the Schr\"odinger equation given in Equation (17). This solution  assumes that $\v B=\gradv\times\v A=0$ holds in the considered region, otherwise the line integral depends on the path. We note that the phase of the wave function can be discontinuous at some point but the wave function must be a continuous function.

Consider the particular case in which  $\v A=\v A_\texttt{string}$, i.e., when the charge $q$ interacts with the potential  $\v A_\texttt{string}$ associated to the string $L'$. With this identification and using cylindrical coordinates, the Dirac condition can be implied by assuming (i) that the path is a closed line surrounding the string
 \begin{align}
\Psi={\rm e}^{i [q/(\hbar c)]\oint_C \v A_\texttt{string}\,\cdot \rho \,d\phi\, \hat{\phi}}\,\Psi_0,
\end{align}
and (ii) that the phase change $[q/(\hbar c)]\oint_C \v A_\texttt{string}\cdot \rho\, d\phi\, \hat{\phi}$ within Equation (69) satisfies the condition
 \begin{align}
\frac{q}{\hbar c}\oint_{C} \v A_\texttt{string}\cdot \rho\, d\phi\, \hat{\phi}=2\pi n.
\end{align}
Under these specific conditions, the possible quantum-mechanical effect of the string on the electric charge will disappear because
${\rm e}^{i [q/(\hbar c)]\oint_C \v A_\texttt{string}\cdot \rho\, d\phi\, \hat{\phi}}={\rm e}^{i 2\pi n}=1$. Integration of the left-hand side of Equation (70) with the potential defined by Equation (60) gives
 \begin{align}
\frac{q}{\hbar c}\oint_C \v A_\texttt{string}\cdot\rho \,d\phi\, \hat{\phi}&=\frac{2qg}{\hbar c}\Theta(\rho-\varepsilon)\Theta(-z)\int\limits_0^{2\pi}d \phi\nonumber\\
&=\frac{4\pi qg}{\hbar c}\Theta(\rho-\varepsilon)\Theta(-z)\nonumber\\
&=\frac{4\pi qg}{\hbar c},
\end{align}
 for $\rho>\varepsilon$ and $z<0$. From Equations (70) and (71), we directly obtain the Dirac quantisation condition $qg=n\hbar c/2$. We then conclude that from quantum-mechanical considerations the unobservability of the string (classically well argued) implies the Dirac condition. The argument can be reversed. If we start by imposing the Dirac condition then the Dirac string turns out to be undetectable.
 The previous treatment to the Dirac string may be seen as a complementary discussion to the heuristic approach to the Dirac condition.
 In the following section, we will review some of the well-known derivations of the Dirac quantisation condition.

\section{Aharonov--Bohm effect and the Dirac quantisation condition}
\label{10}

\noindent We will now review the {\it third quantum-mechanical derivation of the Dirac condition}. According to the Aharonov--Bohm (AB) effect \cite{40}, particles can be affected by a vector potential even in regions where the magnetic field vanishes. We observe that this effect and the derivation of the Dirac quantisation condition require similar objects: a long solenoid for the AB effect and a semi-infinite string for the Dirac condition. Therefore, we may think of the Dirac string as the AB solenoid and investigate as to whether the undetectability of the Dirac string can be demonstrated via a hypothetical AB interference experiment \cite{4,8,9,10,11,12,14,15,17,22,23,41,42}.

\begin{figure}[h]
  \centering
  \includegraphics[width= 210pt]{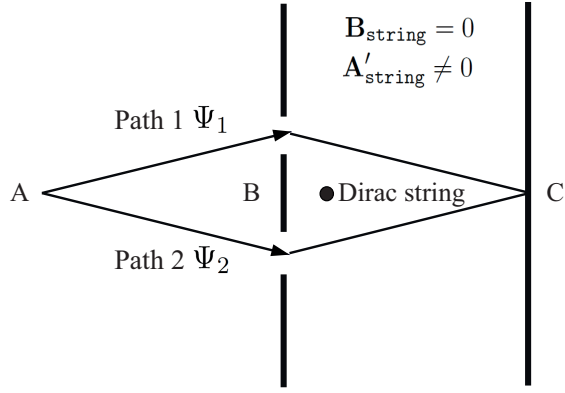}
  \caption{The AB double slit experiment with the Dirac string inserted between the slits. If we demand the string to be undetectable by the wave function it follows that the Dirac quantisation condition holds. Conversely, if the Dirac condition holds then the string is undetectable.}\label{Fig7}
\end{figure}

Let us imagine a double-slit AB experiment with a Dirac string inserted between the slits as shown in Figure \ref{Fig7}. Electric charges are emitted by a source at point A, pass through two slits 1 and 2 of the screen located at point B, and finally are detected at point C. The wave function in a region of zero vector potential is simply $\Psi=\Psi_1\!+ \!\Psi_2 $ where $\Psi_1$  and $\Psi_2$ are the wave functions of the charges passing through the slits 1 and 2. Without the presence of the string, the wave function of the charges combines coherently in such a way that the probability density at C reads $P=|\Psi_1+ \Psi_2|^2.$

Since the Dirac string is inserted between the two slits, it is clear that each of the wave functions $\Psi_1$  and $\Psi_2$ pick up a phase due to the string potential $\v A_\texttt{string}\equiv\v A_s$.  Thus the wave function of the charges is now given by
\begin{align}
\Psi =&\,{\rm e} ^{(iq/\hbar c)\int_{1} \v A_s \cdot \rho\, d\phi\, \hat{\phi}}\,\Psi_1 + {\rm e}^{(iq/\hbar c)\int_{2} \v A_s \cdot \rho \,d\phi\, \hat{\phi}}\,\Psi_2 \nonumber\\=& \, \bigg(\Psi_1 +  {\rm e}^{(iq/\hbar c)\oint_{C} \v A_s \cdot \rho\, d\phi\, \hat{\phi}}\,\Psi_2\bigg) \,{\rm e}^{(iq/\hbar c)\int_{1} \v A_s \cdot \rho\, d\phi\, \hat{\phi}}\nonumber\\
=& \, \bigg(\Psi_1 +  {\rm e}^{i 4\pi q g/(\hbar c)}\,\Psi_2\bigg) \,{\rm e}^{(iq/\hbar c)\int_{1} \v A_s \cdot \rho\, d\phi\, \hat{\phi}},
\end{align}
where we have used the expression for  $\v A_\texttt{string}$ given in Equation (60) and written as
\begin{align}
\oint_{C} \v A_s \cdot \rho\, d\phi\, \hat{\phi}=\int\limits_{2} \v A_s \cdot \rho\, d\phi\, \hat{\phi}- \int\limits_{1} \v A_s \cdot \rho\, d\phi\, \hat{\phi}.
\end{align}
It follows now that the probability density at C reads
\begin{align}
P= |\Psi_1 + {\rm e}^{i4\pi q g/(\hbar c)}\,\Psi_2|^2.
\end{align}
The effect of the Dirac string would be unobservable if ${\rm e}^{i4\pi q g/(\hbar c)}=1$ and this implies the Dirac quantisation condition $qg=n\hbar c/2.$ Under this condition, the probability density becomes $P= |\Psi_1 + \Psi_2|^2$, meaning that no change in the interference pattern would be observed due to the Dirac string. In short: the Dirac string is undetectable if the Dirac quantisation condition holds. We can reverse the argument: if the Dirac quantisation condition holds, then the Dirac string is unobservable.

\section{Feynman's path integral approach and the Dirac quantisation condition}
\label{11}

\noindent
We will now discuss the {\it fourth quantum-mechanical derivation of the Dirac condition}. The path-integral approach to quantum mechanics, suggested by Dirac in 1933 \cite{43},  formally started by Feynman in his 1942 Ph.D. thesis \cite{44}
and fully discussed by him in 1948 \cite{45}, provides an elegant procedure to obtain the Dirac condition, which is similar to a certain extent to that of the Aharonov--Bohm effect.

\begin{figure}[h]
  \centering
  \includegraphics[width=260pt]{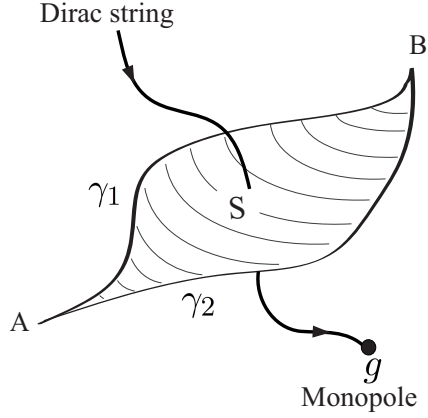}
  \caption{A Dirac string is encircled between two generic paths $\gamma_1$ and $\gamma_2$ starting at A, ending at B, and forming the boundary of the surface ${\rm S}$.}\label{Fig8}
\end{figure}

Let us first briefly discuss the essence of the path-integral approach. Question \cite{59}: If a particle is at an initial position A, what is the probability that it will be at another position B at the latter time? Schr\"odinger's wave function tells us the probability for a particle to be in a certain point in time, but it does not tell us the transition probability for a particle to be between two points at different times. We need to introduce a quantity that generalises the concept of wave function to include transition probabilities. According to Feynman, this concept is the ``transition probability amplitude" (or amplitude for short) which relates the state of a wave function from the initial position and time $\ket{\Psi(\v x_\text{i}, t_\text{i})}$ to its final position and time $\ket{\Psi(\v x_\text{f}, t_\text{f})}$, and is given by the inner product $K=\braket{\Psi(\v x_\text{f}, t_\text{f})|\Psi(\v x_\text{i}, t_\text{i})},$ where we have used Dirac's ``bra-ket" notation. It follows that the transition probability (or probability for short) is defined as $P=|K|^2.$ Dirac \cite{43} suggested that the amplitude for a given path is proportional to the exponent of the classical action associated to the path ${\rm e}^{(i/\hbar) {\cal S}(\v x)},$ where ${\cal S}(\v x)= \int L(\v x, \dot{\v x})dt,$ is the classical action, with $L$ being the Lagrangian. But a particle can take any possible path from the initial to the final point (there is no reason for the particle to take the shortest path). Therefore, to compute the amplitude, Feynman proposed to sum over all the infinite paths that the particle can take. More specifically, the transition probability amplitude $K$ for a charged particle to propagate from an initial point A to a final point B is given by the integral over all possible paths
\begin{align}
K=\int \!\mathcal{D(\v x)}\,{\rm e}^{(i/\hbar) {\cal S}(\v x)},
\end{align}
where $\int\mathcal{D(\v x)}$ is a short hand to indicate a product of integrals performed over all paths $\v x(t)$ leading from $\text{A}$ to $\text{B},$  and ${\cal S}$ is the classical action associated to each path. For example, consider two generic paths $\gamma_1$ and $\gamma_2$ each of which starts at $\text{A}$ and ends at $\text{B}.$  The amplitude is
\begin{align}
K=K_1+K_2=\int\limits_{\gamma_1} \!\mathcal{D(\v x)}\, {\rm e}^{(i/\hbar) {\cal S}^{(1)}(\v x)} + \int\limits_{\gamma_2}\!\mathcal{D(\v x)}\,{\rm e}^{(i/\hbar) {\cal S}^{(2)}(\v x)},
\end{align}
where $K_1$ is the amplitude associated to the integration over all paths through $\gamma_1$ and $K_2$ is the amplitude associated to the integration over all paths through $\gamma_2.$ Consider first the action for a free particle ${\cal S}_0= \int \! m\dot{\v x}^2/2\,dt.$ In this case, there is not external interaction and therefore the probability is simply $P=|K_1+K_2|^2.$  Nothing really interesting happens there. Consider now the case where the electric charge is affected by the potential due to the magnetic monopole and the Dirac string given in Equation (2). Furthermore, suppose that the paths $\gamma_1$ and $\gamma_2$ pass on each side of the Dirac string and form the boundary of a surface ${\rm S}$ as seen in Figure \ref{Fig8}.
\\
\\
\\

The external vector potential $\v A_L$ will affect the motion of the particle because the action acquires an interaction term
\begin{align}
{\cal S}= {\cal S}_0 + \frac{q}{c} \int  \v A_L \cdot d \v l.
\end{align}
Thus the amplitude becomes
\begin{align}
\nonumber K=& \int\limits_{\gamma_1} \!\mathcal{D}(\v x)\,{\rm e}^{(i/\hbar) ({\cal S}^{(1)}_{0}+ (q/c)\int_{(1)}  \v A_L \cdot d \v l)} + \int\limits_{\gamma_2} \!\mathcal{D}(\v x)\,{\rm e}^{(i/\hbar) ({\cal S}^{(2)}_{0} + (q/c)\int_{(2)}  \v A_L \cdot d \v l)}
\\ =& \bigg(K_1 + {\rm e}^{ (i q/\hbar c)\oint_{C} \v A_L \cdot d \v l }\,K_2\bigg) {\rm e}^{ (i q/\hbar c)\int_{(1)}  \v A_L \cdot d \v l},
\end{align}
where we have written
\begin{align}
\oint_{C} \v A_L \cdot d \v l=\int_{(2)}  \v A_L \cdot d \v l- \int_{(1)}  \v A_L \cdot d \v l.
\end{align}
Clearly, the contributions from $\gamma_1$ and $\gamma_2$ interfere, giving the interference term ${\rm e}^{(i q/\hbar c)\oint_{C} \v A_L \cdot d \v l }.$ Using Stoke's theorem and Equation (4) we can write the integral of this exponent as
\begin{align}
 \oint_{C} \v A_L \cdot d \v l = \int_{S} \gradv \times \v A_L \cdot d\v a = \int_{\rm S} \v B_\texttt{mon} \cdot d \v a + \int_{S} \v B_\texttt{string} \cdot d\v a.
\end{align}
Therefore, we may write the interference term as
\begin{align}
{\rm e}^{(i q/\hbar c)\oint_{C} \v A_L \cdot d \v l }= {\rm e}^{(iq/\hbar c) \int_{\rm S} \v B_\texttt{mon} \cdot d \v a} \, {\rm e}^{ (iq/\hbar c)\int_{\rm S} \v B_\texttt{string} \cdot d\v a}.
\end{align}
The term ${\rm e}^{(iq/\hbar c) \int_{\rm S} \v B_\texttt{mon} \cdot d \v a}$  is perfectly fine because the charged particle should be influenced by the magnetic monopole. However, the second term must not contribute or otherwise the string would be observable. Therefore, we must demand ${\rm e}^{ (iq/\hbar c)\int_{\rm S} \v B_\texttt{string} \cdot d\v a}=1.$ But the flux through the string is $\int_{\rm S} \v B_\texttt{string} \cdot d\v a=4 \pi g$ so that ${\rm e}^{i4 \pi qg/\hbar c} =1,$ which implies the Dirac quantisation condition $qg= n \hbar c /2.$

As may be seen, the procedure to obtain the Dirac quantisation condition based on Feynman's path integral approach is similar to the procedure based on the Aharonov--Bohm effect. If one first teaches the latter procedure in an advanced undergraduate course, then one may teach the former procedure in a graduate course, following  Feynman's opinion that \cite{45}: ``there is a pleasure in recognising old things from a new point of view."

\section{The Wu--Yang approach and the Dirac quantisation condition}
\label{12}

\noindent We will now examine the {\it fifth quantum-mechanical derivation of the Dirac condition}. Let us rewrite Equations (40) and (41) as follows:
\begin{align}
\v B'\!&=\gradv \times\v A'= \;\frac{g}{r^2}\hat{\v r} +  4\pi g\delta(x)\delta(y)\Theta(-z)\hat{\v z},\\
\v B &=\gradv \times\v A = \;\frac{g}{r^2}\hat{\v r} - 4\pi g\delta(x)\delta(y)\Theta(z)\hat{\v z}.
\end{align}
A direct look at these equations reveals an unpleasant but formal result: $\v B'\not=\v B.$
This result follows from the difference of the delta-field contributions of the respective strings. Therefore, the potentials $\v A'$ and $\v A$ are not equivalent. Strictly speaking they are not gauge potentials. However, it is possible to extend the gauge symmetry to include contributions due to strings \cite{8}, but this possibility will not be discussed here. Using the property $\Theta(-z)=1-\Theta(z)$, the difference of the magnetic fields is given by $\v B'-\v B= 4\pi g\delta(x)\delta(y)\hat{\v z},$ where the right-hand side of this equation  is a singular magnetic field attributable to an infinite string lying along the entire $z$-axis.
The fact that $\v B'$ and $\v B$ are different is not an unexpected result because the current densities producing them are different as may be seen in Equations (50) and (52). However, we have argued that the Dirac strings are unphysical and should therefore be unobservable. The question then arises: How should the potentials $\v A'$ and $\v A$ be interpreted? A rough answer will be that $\v A'$ and $\v A$ are equivalent because they produce the same magnetic field [the first terms of Equations (82) and (83)] and because the field contributions of the strings [the last terms of Equations (82) and (83)] can be physically ignored. But we must recognise that this answer is not very satisfactory from a formal point of view. In other words,  $\v A'$ and $\v A$ are physically but not mathematically equivalent.

Furthermore, it can be argued that the derivation of the Dirac condition involves some unpleasant features like singular gauge transformations and singular potentials \cite{9}. Fortunately, a procedure due to Wu and Yang \cite{46} avoids these unpleasant features and leads also to the Dirac condition. The Wu--Yang method does not to deal with singular potentials nor with singular gauge transformations (except with the real singularity at the origin). The strategy of  Wu and Yang was to use different vector potentials in different regions of space. In more colloquial words, if the Dirac string is the cause of the difficulties and subtleties, then the Wu-Yang approach provides a simple solution: to get rid of the Dirac string via a formal procedure.

\begin{figure}[h]
  \centering
  \includegraphics[width=228pt]{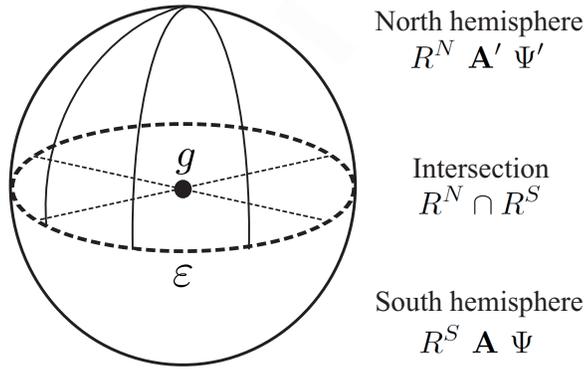}
  \caption{The Wu-Yang configuration describing a magnetic monopole without the Dirac strings.}\label{Fig9}
\end{figure}

In the Wu--Yang method the potentials $\v A'$ and $\v A$ displayed in Equation (32) are non-singular if we define them in an appropriate domain:
\begin{align}
\v A'=& \,\,g\frac{1-\cos\theta}{r \sin\theta}\hat{\phi},\qquad\quad R^N:\;0 \leq \theta < \frac{\pi}{2}+\frac\varepsilon2\\
\v A =&  -g\frac{1+\cos\theta}{r \sin\theta}\hat{\phi},\quad \quad R^S:\;\frac{\pi}{2}-\frac\varepsilon2 < \theta \leq \pi
\end{align}
where $\varepsilon>0$ is an infinitesimal quantity. The potentials  $\v A'$ and $\v A$ are in the Coulomb gauge: $\gradv \cdot \v A =0$ and $\gradv \cdot \v A'=0$. Furthermore, these potentials are non-global functions since they are defined only on their respective domains: $R^N$ and $R^S$. The region $R^N$, where $\v A'$ is defined, excludes the string along the negative semi-axis $(\theta=\pi)$ and represents a North hemisphere. The region $R^S$, where $\v A$ is defined, excludes the string along the positive semi-axis $(\theta=0)$ and represents a South hemisphere. The union of the  hemispheres $R^N\cup R^S$ covers the whole space (except on the origin, where there is a magnetic monopole). In the intersection $R^N\cap R^S$  (the ``equator'') both hemispheres are slightly overlapped. A representation of the  Wu-Yang configuration is shown in Figure \ref{Fig9}.

Using Equation (28), the potentials $\v A'$ and $\v A$ defined by Equations (84) and (85) yield the field of a magnetic monopole: $\v B = \gradv \times \v A'= \gradv \times \v A= g \hat{\v r}/r^2$. Therefore, the potentials  $\v A'$ and $\v A$ must be connected by a gauge transformation in the overlapped region $\pi/2 - \varepsilon/2 < \theta <\pi/2 + \varepsilon/2$, where both potentials are well defined. At first glance, $\v A'-\v A=2 g\hat{\phi}/(r \sin\theta)$. But in the overlapped region, we have $\lim \sin(\pi/2\pm\varepsilon/2)=1$ as $\varepsilon\to 0$ and thus
\begin{align}
\v A'-\v A=\frac{2 g}{r}\hat{\phi}=\gradv (2g\phi)=\gradv \Lambda,
\end{align}
where $\Lambda =2 g \phi$ (the gauge function $\Lambda$ satisfies $\gradv^2\Lambda=0$ indicating that $\v A'$ and $\v A$ are related by a restricted gauge transformation).  Suppose now that an electric charge is in the vicinity of the magnetic monopole. In this case, we require two wave functions to describe the electric charge: $\Psi'$ for $ R^N$ and $\Psi$ for $ R^S$. In the overlapped region, the wave functions $\Psi'$ and $\Psi$ must be related by the phase transformation $\Psi'={\rm e}^{iq\Lambda/(\hbar c)}\,\Psi$, which is associated to the gauge transformation given in Equation (86). This phase transformation with $\Lambda =2 g \phi$ reads
\begin{align}
\Psi'={\rm e}^{i2qg \phi/(\hbar c)}\,\Psi.
\end{align}
But the wave functions $\Psi'$ and $\Psi$ must be single-valued \big($\Psi'|_{\phi} = \Psi'|_{\phi + 2\pi}\big),$ which requires ${\rm e}^{i4 \pi qg/ (\hbar c )}\!=\!1,$ and this implies the Dirac quantisation condition $qg=n\hbar c/2$. Remarkably, Equations (84)-(87) do not involve unpleasant singularities.  The approach suggested by Wu and Yang constitutes a refinement of Dirac's original approach. It is pertinent to say that the Wu--Yang approach has become popular in many treatments of the Dirac quantisation condition \cite{4,8,9,11,12,13,16,24}.

\section{Semi-classical derivations of the Dirac quantisation condition}
\label{13}

\noindent We will now discuss the {\it first semi-classical derivation of the Dirac condition}. In 1936, Saha wrote \cite{33}: ``If we take a point charge $e$ at A and a magnetic pole $\mu$ at B, classical electrodynamics tells us that the angular momentum of the system about the line AB is just $e\mu/c$. Hence, following the quantum logic, if  we put this $= h/(2\pi)$, the fundamental unit of angular momentum, we have $\mu= ch/(4\pi e)$ which is just the result obtained by Dirac.'' This relatively simple semi-classical argument to arrive at the Dirac condition [with $n=1$] remained almost ignored until 1949 when Wilson \cite{36,37} used the same argument to obtain this condition [now with $n$ integer]. Let us develop in more detail the derivation of Dirac's condition suggested by Saha and also by Wilson. When the Dirac condition is written as $qg/c=n\hbar/2,$ we can see that the left-hand side has units of angular momentum because the constant $\hbar$ has these units. This suggests the possibility that the quantity $qg/c$ can be obtained from the electromagnetic angular momentum:
\begin{align}
\v L_\texttt{EM} = \frac{1}{4 \pi c} \int_{V} \v x \times (\v E \times \v B)\,d^3x,
\end{align}
with the idea that the field $\v E$ is produced by the electric charge $q$ and the field $\v B$ by the magnetic charge $g$, both charges at rest and separated by a finite distance. This configuration was considered by Thomson \cite{34,35} in 1904, and is now known as the ``Thomson dipole.'' More precisely stated, the Thomson dipole  is a static dipole formed by an electric charge $q$ and a magnetic charge $g$ separated by the distance $a\!=\!|\v a|,$ where the vector $\v a$ is directed from the charge $q$ to the charge $g.$ For convenience, we place the charge $q$ at $\v x'=-{\v a}/2$ and the charge $g$ at $\v x'\!=\!{\v a}/2$ as seen in Figure \ref{Fig10}. Clearly, there is no mechanical momentum associated to this dipole because it is at rest. In Appendix \ref{E}, we show that the electromagnetic angular momentum due to the fields of the charges $q$ and $g$ is given by
\begin{align}
\v L_\texttt{EM}= \frac{qg}{c}\hat{\v a},
\end{align}
where $\hat{\v a}=\v a/a$. This equation was derived by Thomson \cite{34,35}. Remarkably, the magnitude of $\v L_\texttt{EM}$ does not depend on the distance between the charges. We note that Equation (89) has been derived by several equivalent procedures \cite{60,61}. Notice also that this angular momentum is conserved: $d \v L_\texttt{EM}/dt = 0.$ We now invoke a quantum mechanical argument: quantisation of the angular momentum. As is well known in quantum mechanics, the total (conserved) angular momentum operator $\widehat{\cJ}$ of a system reads \cite{21}: $\widehat{\cJ}= \widehat{\cL} + \widehat{\cS},$ where $\widehat{\cL}$ is the orbital angular momentum operator and $\widehat{\cS}$ is the spin angular momentum operator. In order to obtain
$\widehat{\cJ}$ for a given system, we first identify its corresponding classical counterpart.
\begin{figure}[h]
  \centering
  \includegraphics[width= 248pt]{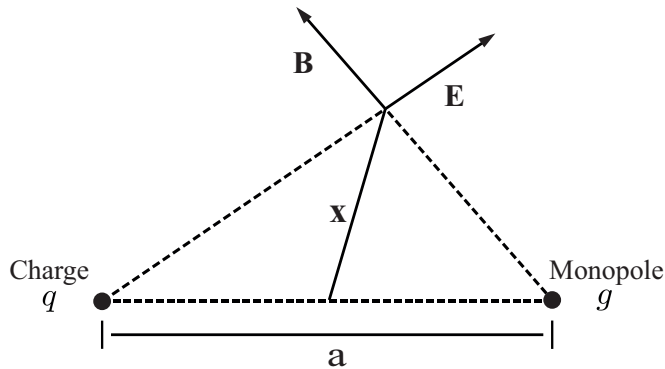}
  \caption{Configuration of the Thomson dipole.}\label{Fig10}
\end{figure}
Evidently, the Thomson dipole lacks of an orbital angular momentum. We can therefore identify $\widehat{\cS}$ with $\widehat{\cJ}$ and make the substitution $\v L_\texttt{EM}\rightarrow \widehat{\cJ}.$ If we measure $\widehat{\cJ}$ along any of its three spatial components, say $z,$ it takes the discrete values $J_z = n\hbar/2$ \cite{21}. Therefore, if we choose $\hat{\v a} = \hat{\v z}$ in Equation (89) then we can quantise the $z$ component of this equation. Following this argument we obtain $J_z = qg/c =n\hbar/2,$ which yields the Dirac condition $qg=n \hbar c /2.$ We should emphasise that this method is semiclassical in the sense that the angular momentum $qg/c$ is first obtained from purely classical considerations and then it is made equal to $n\hbar/2$ by invoking a quantum argument.

\begin{figure}[h]
  \centering
  \includegraphics[width= 160pt]{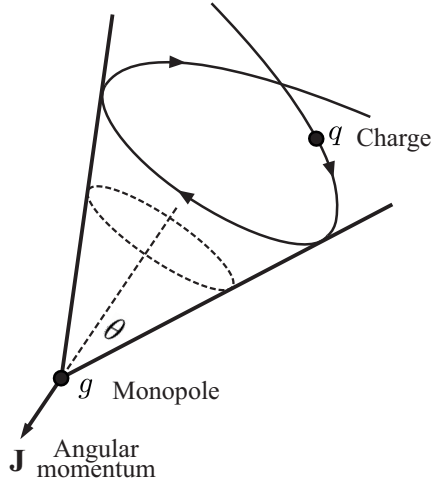}
  \caption{Dynamics of a moving electric charge in the field of a magnetic monopole. In this configuration the angular momentum $\hat{\v r}\cdot \v J = -qg/c$ is constant. This means that the charge moves in a cone on the axis $\v J,$ with the angle $\theta = \cos^{-1}(qg/Jc).$}\label{Fig11}
\end{figure}

We will now examine the {\it second semi-classical  derivation of the Dirac condition}.
We can also arrive at the Dirac condition by another semiclassical method due to Fierz \cite{38}. Consider an electric charge $q$ moving with velocity $\dot{\v x}$ in the field of a monopole $g$ centred at the origin: $\v B= g\hat{\v r}/r^2.$ This configuration is illustrated in Figure \ref{Fig11}.
The charge $q$ experiences the Lorentz force
\begin{align}
\frac{d\v p}{dt}  = q\bigg(\frac{\dot{\v x}}{c}\times \v B\bigg),
\end{align}
where $\v p = m \dot{\v x}$ is the mechanical momentum associated to the charge $q$. The field of the monopole is spherically symmetric and therefore one should expect the total angular momentum of the system is conserved. To see this, we take the cross product of Equation (90) with the position vector $\v x,$ use $\v x \times (d\v p/dt) = d(\v x\times \v p)/dt,$ and obtain the corresponding torque
\begin{align}
 \frac{d (\v x\times \v p)}{dt}=&\, \frac{q}{c}\big(\v x \times (\dot{\v x}\times \v B)\big)= \frac{qg}{c}\bigg( \frac{\v x \times (\dot{\v x}\times \v x)}{r^3}\bigg)= \frac{d}{dt}\bigg(\frac{qg }{c}\hat{\v r} \bigg),
\end{align}
where we have used the identity
\begin{align}
\frac{\v x \times (\dot{\v x}\times \v x)}{r^3} =\frac{d \hat{\v r}}{dt}.
\end{align}
Clearly, the mechanical angular momentum $\v x \times \v p$ is not conserved $d(\v x \times \v p)/dt \neq 0.$ This is an expected result because there is an extra contribution attributed to the angular momentum of the electromagnetic field. From Equation (91), it follows
\begin{align}
\frac{d }{dt} \bigg(\v x\times \v p - \frac{qg }{c}\hat{\v r} \bigg)=0.
\end{align}
Hence, the total (conserved) angular momentum is
\begin{align}
\v J = \v x \times \v p - \frac{qg }{c}\hat{\v r}.
\end{align}
This interesting result was observed by Poincar\'e \cite{62} in 1896, although it was already anticipated by Darboux in 1878 \cite{63}. From Equation (94), it follows that the radial component of this angular momentum is constant $\v J \cdot \hat{\v r} = - qg/c.$ With regard to the quantity  $qg/c$, Fierz \cite{38} pointed out: ``...the classic value $qg/c,$ must be in quantum theory equal to an integer or half-integer multiple of $\hbar.$'' Following this argument, we can quantise the radial component of the angular momentum in Equation (94): $J_r=qg/c=n \hbar/2$ (the minus sign is absorbed by $n$) and this yields the Dirac condition $qg=n\hbar c/2.$

\begin{figure}[h]
 \centering
  \includegraphics[width= 148pt]{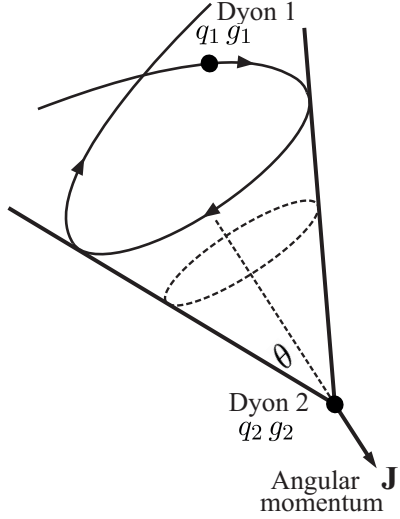}
  \caption{Dynamics of a moving dyon in the field of a stationary dyon. In this configuration the angular momentum $\hat{\v r}\cdot \v J = -(q_1g_2-q_2g_1)/c$ is constant. This means that the dyon moves in a cone on the axis $\v J,$ with the angle $\theta = \cos^{-1}((q_1g_2-q_2g_1)/Jc).$}\label{Fig12}
\end{figure}

We will now review the {\it third semi-classical derivation of the Dirac condition}. Strictly speaking, we will review the derivation of a generalised duality-invariant form of this condition due to Schwinger \cite{39}. The approach followed by Schwinger is similar to that of Fierz but now applied to the case of dyons, which are particles with both electric and magnetic charge. The approach considers the interaction of a dyon of mass $m$ carrying an electric charge $q_1$ and a magnetic charge $g_1,$ moving with velocity $\dot{\v x}$ in the field of a stationary dyon with electric charge $q_2$ and magnetic charge $g_2$ centred at the origin, as seen in Figure \ref{Fig12}. The Lorentz force due to the moving dyon takes the duality-invariant form
\begin{align}
\frac{d\v p}{dt}  = q_1\bigg(\v E + \frac{\dot{\v x}}{c}\times \v B\bigg)+ g_1 \bigg( \v B - \frac{\dot{\v x}}{c}\times \v E \bigg),
\end{align}
where the electric and magnetic fields produced by the charges $q_2$ and $g_2$ of the stationary dyon are
\begin{align}
\v E =\frac{q_2}{r^2}\hat{\v r}, \quad \v B = \frac{g_2}{r^2}\hat{\v r}.
\end{align}
Therefore, we may write Equation (95) as
\begin{align}
\frac{d\v p}{dt} =\big( q_1 q_2 + g_1 g_2\big)\frac{\hat{\v r}}{r^2} + \big(q_1g_2-q_2g_1 \big)\frac{\dot{\v x} \times \v x}{c \,r^3}.
\end{align}
To find the conserved angular momentum of the system, we take the cross product of Equation (97) with the position vector $\v x$, use $\v x \times (d\v p/dt) = d(\v x\times \v p)/dt,$ and obtain
\begin{align}
\frac{d (\v x\times \v p)}{dt}=  \, \frac{\big(q_1g_2-q_2g_1 \big)}{c}\,\frac{d \hat{\v r} }{dt},
\end{align}
where we have used Equation (92). The conserved angular momentum is thus
\begin{align}
\v J= \v x\times \v p - \big(q_1g_2-q_2g_1 \big)\frac{\hat{\v r}}{c},
\end{align}
whose radial component $ \v J \cdot \hat{\v r}= -(q_1g_2-q_2g_1)/c$ can be quantised: $J_r=(q_1g_2-q_2g_1)/c=n \hbar/2,$ yielding the Schwinger--Swanziger quantisation condition
\begin{align}
q_1g_2-q_2g_1 = \frac{n}{2}\hbar c.
\end{align}
In contrast to the Dirac condition $qg=n\hbar c/2$, which for a fixed value of $n$ is not invariant under the dual changes $q\to g$ and $g\to -q$, the Schwinger--Swanziger condition is clearly invariant under these dual changes. Equation (100) was first obtained by Schwinger \cite{64} and independently by Swanziger \cite{65}. Interestingly, both of these authors argued that the quantisation in Equation (100) should take integer and not half-integer values, i.e. Equation (100) should be written as $q_1g_2-q_2g_1 = n \hbar c.$

\section{Final remarks on the Dirac quantisation condition}
\label{14}

The advent of the Dirac quantisation condition brought us two news: one good and another bad. The good news is that this condition allows us to explain the observed quantisation of the electric charge. The bad news is that such an explanation is based on the existence of unobserved magnetic monopoles. One is left with the feeling that the undetectability of magnetic monopoles spoils the Dirac quantisation condition. Evidently, the fact that the Dirac condition explains the electric charge quantisation cannot be considered as a proof of the existence of magnetic monopoles.
Although it has recently been argued that magnetic monopoles may exist, not as elementary particles, but as emergent particles (quasiparticles) in exotic condensed matter magnetic systems such as ``spin ice'' \cite{66,67,68}, there is still no direct experimental evidence of Dirac monopoles. However, experimental searches for monopoles continue to be of great interest \cite{69,70,71,72,73,76}. It can be argued that the idea of undetected magnetic monopoles is too high a price to pay for explaining the observed charge quantisation. But equally it can be argued that magnetic  monopoles constitute an attractive theoretical concept, which is not precluded by any fundamental theory and has been extremely useful in modern gauge field theories \cite{4,29}.

In any case, magnetic monopoles are like the Loch Ness monster, much talked about but never seen. Although many theoretical physicists would say that the idea of magnetic monopoles is too attractive to set aside, we think it would be desirable to have a convincing explanation for the electric charge quantisation without appealing to magnetic monopoles.

It is interesting to note that the introduction of magnetic monopoles in Dirac's 1931 paper \cite{2} was not taken fondly by Dirac himself. He wrote: ``The theory leads to a connection, namely, $[eg_0=\hbar c/2]$, between the quantum of magnetic pole and the electronic charge. It is rather disappointing to find this reciprocity between electricity and magnetism, instead of a purely electronic quantum condition such as [$\hbar c/e^2$].''
However, no satisfactory explanation for the charge quantisation was proposed between 1931 and 1948 and this seemed to led him to reinforce his idea about magnetic monopoles. In his 1948 paper he wrote \cite{3}: ``The quantisation of electricity is one of the most fundamental and striking features of atomic physics, and there seems to be no explanation for it apart from the theory of poles. This provides some grounds for believing in the existence of these poles.''

The story of the Dirac quantisation condition may be traced to the story of a man [P. A. M. Dirac: the theorist of theorists!]
who wanted to know why the electric charge is quantised and why the electric charge of the electron
had just the numerical value that makes the inverse of the fine structure constant to acquire the value $\alpha^{-1}=\hbar c/e^2\approx 137$. Many years later, he expressed his frustration at not being able to find this magic number. He criticised his theory because it \cite{30}: ``...did not lead to any value for this number $[\alpha^{-1}\approx 137],$ and, for that reason, my argument seemed to be a
failure and I was disappointed with it.'' But the idea of explaining this number seems to have been always important for him. With the confidence of a master, Dirac wrote  \cite{30}: ``The problem of explaining this number $\hbar c/e^2$ is still completely unsolved. Nearly 50 years have passed since then. I think it is perhaps the most fundamental unsolved problem of physics at the present time, and I doubt very much whether any really big progress will be made in understanding the fundamentals of physics until it is solved.''

Although Dirac was not successful in explaining why the charge of the electron has its observed value, in the search for this ambitious goal, he envisioned a magnetic monopole attached to a semi-infinite string, which he required to be unobservable by a quantum argument, obtaining thus a condition that explains the electric charge quantisation. This is indeed a brilliant idea not attributable to an ordinary genius but rather to a magician, a person ``whose inventions are so astounding, so counter to all the intuitions of their colleagues, that it is hard to see how any human
could have imagined them'' \cite{74}.

\section{A final comment on nodal lines}
\label{15}
Berry \cite{77} has pointed out that the nodal lines introduced by Dirac in his 1931 paper \cite{2} are an example of dislocations in the probability waves of quantum mechanics. The history can be traced to 1974 when Nye and Berry \cite{78} observed that wavefronts can contain dislocation lines, closely analogous to those found in crystals. They defined these dislocation lines as those lines on which the phase of the complex wave function is
undetermined, which requires the amplitude be zero, indicating that dislocation lines are lines of singularity (or lines of zeros). Remarkably, the
 lines of singularity (also called wave dislocations, nodal lines, phase singularities and wave vortices) are generic features of waves of all kinds, such as light waves, sound waves and quantum mechanical waves. These lines involve two essential properties: on these lines the phase is singular (undetermined) and around these lines the phase changes by a multiple (typically $\pm1$) of $2\pi.$ Even though the concept of the line of singularity has been extensively discussed in the literature (see, for example, the collection of papers in the special issues mentioned in References \cite{79,80,81,82}), its connection with the Dirac strings is not usually commented on. In his review on singularities in waves \cite{77}, Berry has claimed: ``He [Dirac] recognises that $\Psi_0$ [appearing in Equation (68)] can have nodal lines around which the phase $\chi_0$ in the absence of magnetic field changes by $2n\pi$, i.e. he recognises the existence of wavefront dislocations.'' However, it should be emphasised that the semi-infinite nodal lines introduced by Dirac are unobservable because of the Dirac quantisation condition. But in the general case, the lines of singularity are physical and can form closed loops, which can be linked and knotted \cite{83}.

\section{Conclusion}
\label{16}

\noindent In this review paper, we have discussed five quantum-mechanical derivations, three semiclassical derivations and a novel heuristic derivation of the Dirac quantisation condition. They are briefly resumed as follows.
\vskip 5pt
 {\it First quantum mechanical derivation}. In this derivation, the magnetic monopole is attached to an infinite line of dipoles, the so-called Dirac string \cite{18}. The vector potential of this configuration yields the field of the magnetic monopole plus a singular magnetic field due to the Dirac string. By assuming that the location of the string must be irrelevant, it is shown that the two arbitrary positions of the string are connected with two gauge potentials, meaning that the change of a string to another string is equivalent to a gauge transformation involving a multi-valued gauge function. By demanding the wave function in the phase transformation
be single-valued, the Dirac condition is required.
\vskip 5pt
{\it Heuristic derivation.} (i) It starts with the relation ${\rm e}^{i 2k\phi}={\rm e}^{iq \Lambda/(\hbar c)}$, where $k$ is an arbitrary constant, $\phi$ the azimuthal angle and $\Lambda$ an unspecified gauge function; (ii) from this relation it follows the remarkable equation
$\Lambda q/(\hbar c)=2 k \phi$. One solution of this equation is given by $k=qg/(\hbar c)$ and $\Lambda= 2g \phi$, where $g$ is a constant to be identified; (iii) if the phase ${\rm e}^{iq \Lambda/(\hbar c)}$ is required to be single-valued, then ${\rm e}^{i 2k\phi}$ must be also single-valued and this implies the ``quantisation" condition $k=n/2$ with $n$ being an integer; (iv) from this condition and $k=qg/(\hbar c),$ we get the relation $qg=n\hbar c/2$; (v) the function  $\Lambda= 2g \phi$ with $g$ being the magnetic charge is proved to be a gauge function and this allows us to finally identify $q g=n\hbar c/2$ with the Dirac quantisation condition; (vi) a weak point of this heuristic derivation is that the associated Dirac strings are excluded; (vii) classical considerations indicate that the Dirac string lacks of physical meaning and is thus unobservable; (viii) Quantum mechanical considerations show that the undetectability of the Dirac string implies the Dirac condition.
\vskip 5pt
{\it Second  quantum mechanical derivation}.
The quantum-mechanical interaction of an electric charge $q$ with the potential $\v A$ is given by the phase appearing in the wave function $\Psi={\rm e}^{i[q/(\hbar c)]\int_0^{\v x}\v A(\v x')\cdot d\v l'}\Psi_0,$ where $\Psi_0$ is the solution of the free Schr\"odinger equation and the line integral in the phase is taken a long a path followed by $q$ from the origin to the point $\v x$. If $\v A=\v A_\texttt{string}=2g\Theta(\rho-\varepsilon)\Theta(-z)\hat{\phi}/\rho$ and the path is a closed line surrounding  the string, we have $[q/(\hbar c)]\oint_C \v A_\texttt{string}\cdot \rho \,d\phi\, \hat{\phi}= 4\pi qg/(\hbar c)$ for $\rho>\varepsilon$ and $z<0$.  If now we demand this quantity to be equal to $2\pi n,$ then the effect of the string on the charge $q$ disappears because ${\rm e}^{i 4\pi qg/(\hbar c)}={\rm e}^{i 2\pi n}=1$ and this implies the Dirac condition.
\vskip 5pt

{\it Third quantum mechanical derivation.} This derivation is directly related to the Aharonov--Bohm double-slit experiment \cite{40} with the Dirac string  inserted between the slits. Considering the vector potential of the string, it is shown that the corresponding probability density is $P= |\Psi_1 + {\rm e}^{i4\pi q g/(\hbar c)}\,\Psi_2|^2.$
The effect of the Dirac string is unobservable if ${\rm e}^{i4\pi q g/(\hbar c)}=1$ and this implies the Dirac condition. Vice versa, if this condition holds \emph{a priori} then the Dirac string is unobservable.
\vskip 5pt
{\it Fourth quantum mechanical derivation.}
According to Feynman's path-integral approach to quantum mechanics \cite{45}, the amplitude of a particle reads $K=\int \!\mathcal{D(\v x)}\,{\rm e}^{(i/\hbar) {\cal S}(\v x)}$, where $\int\mathcal{D(\v x)}$ indicates a product of integrals performed over all paths $\v x(t)$ going from $\text{A}$ to $\text{B},$  and ${\cal S}$ is the classical action associated to each path. For two such generic paths in free space, $\gamma_1$ and $\gamma_2$, we have $K=K_1+K_2=\!\int_{\gamma_1} \!\!\mathcal{D(\v x)}\, {\rm e}^{(i/\hbar){\cal S}^{(1)}(\v x)} \!+\! \int_{\gamma_2}\!\!\mathcal{D(\v x)}\,{\rm e}^{(i/\hbar) {\cal S}^{(2)}(\v x)}.$ Suppose that $\gamma_1$ and $\gamma_2$ pass on each side of the Dirac string and form the boundary of a surface S. As a result, the action acquires an interaction term ${\cal S}\!=\! {\cal S}_0 + (q/c)\int  \v A_L \cdot d \v l,$ where ${\cal S}_0$ is the action for the free path. Thus the amplitude becomes $K\!=\! \big(K_1\! + \!{\rm e}^{(i q/\hbar c)\oint_{C} \v A_L \cdot d \v l }{ K_2}\big) {\rm e}^{ (i q/\hbar c)\int_{(1)}  \v A_L \cdot d \v l},$ and the interference term is ${\rm e}^{(i q/\hbar c)\oint_{C} \v A_L \cdot d \v l }.$ Using the Stoke's theorem and $\gradv\times \v A_L\! =\!\v B_\texttt{mon}+\v B_\texttt{string}$, the interference term becomes ${\rm e}^{(i q/\hbar c)\oint_{C} \v A_L \cdot d \v l }\!=\! {\rm e}^{(iq/\hbar c) \int_{s} \v B_\texttt{mon} \cdot d \v a} \, {\rm e}^{ (iq/\hbar c)\int_{\rm S} \v B_\texttt{string} \cdot d\v a}.$ The second exponential factor on the right should not contribute or otherwise the string would be observable. Thus we must demand ${\rm e}^{ (iq/\hbar c)\int_{S} \v B_\texttt{string} \cdot d\v a}=1.$ But the flux through the string is $\int_{\rm S} \v B_\texttt{string} \cdot d\v a=4 \pi g$ so that ${\rm e}^{i 4 \pi qg/\hbar c} =1,$ which implies Dirac's condition.

\vskip 5pt
{\it Fifth quantum mechanical derivation.} This derivation describes a magnetic monopole without Dirac strings \cite{46} using two non-singular potentials which are defined in two different regions of space. In the intersection region, both potentials are connected by a non-singular gauge transformation with the gauge function $\Lambda =2 g \phi$. The description of an electric charge in the vicinity of the magnetic monopole requires two wave functions $\Psi'$ and $\Psi$, which are related by the phase transformation $\Psi'={\rm e}^{i2qg \phi/(\hbar c)}\Psi$ in the overlapped region. But $\Psi'$ and $\Psi$ must be single-valued \big($\Psi'|_{\phi}\!=\!\Psi'|_{\phi+2\pi}\big),$ which requires ${\rm e}^{i4 \pi qg/ (\hbar c )}\!=\!1,$ and this implies Dirac's condition.

\vskip 5pt
{\it First semi-classical derivation.} This derivation considers the Thomson dipole \cite{34,35}, which is a static dipole formed by an electric charge $q$ and a magnetic charge $g$ separated by the distance $a\!=\!|\v a|$ \cite{60,61}. The electromagnetic angular momentum of this dipole is given by $\v L_\texttt{EM}= qg\hat{\v a}/c.$ By assuming that any of the spatial components of the angular momentum must be quantised in inter multiples of $\hbar/2$, we obtain Dirac's condition.
\vskip 5pt
{\it Second semi-classical derivation.} This derivation considers an electric charge $q$ moving with speed $\dot{\v x}$ in the field of a monopole $g$ \cite{8,38}. The associated Lorentz force $d\v p/dt= q\big(\dot{\v x}\times \v B/c\big)$ is used to obtain total (conserved) angular momentum of this system
$\v J = \v x \times \v p - qg\hat{\v r}/c.$ The radial component $ \v J \cdot \hat{\v r}= -qg/c$ is then quantised yielding Dirac's condition.

\vskip 5pt
{\it Third semi-classical derivation.} This derivation considers a dyon of mass $m$ carrying an electric charge $q_1$ and a magnetic charge $g_1,$ moving with velocity $\dot{\v x}$ in the field of a stationary dyon with charge $q_2$ and $g_2$ located at the origin \cite{39}. Using the duality-invariant form of the Lorentz force $d\v p/dt  = q_1\big(\v E + \dot{\v x}\times \v B/c\big)\!+\! g_1 \big( \v B -\dot{\v x}\times \v E/c \big)$ the total angular momentum of this system is found to be $\v J= \v x\times \v p - \big(q_1g_2-q_2g_1 \big)\hat{\v r}/c.$ The radial component  $ \v J \cdot \hat{\v r}= -(q_1g_2-q_2g_1)/c$ is then quantised yielding the Schwinger--Swanziger condition $q_1g_2-q_2g_1 = n\hbar c/2$ which is a duality invariant form of
Dirac's condition.

\section*{Note}
\noindent A derivation of Equations (17)-(19), which is more pedagogical than that appearing in the standard graduate textbooks (for example in Reference \cite{21}), is available in the author's website: \href{http://ricardoheras.com/}{www.ricardoheras.com}.

\section*{Acknowledgements}
I wish to thank Professor Michael V. Berry for bringing my attention to the important topic of wavefront dislocations and its connection with the Dirac strings.

\section*{Notes on contributor}
\begin{minipage}{0.2\textwidth}
\scalebox{0.42}{\includegraphics[origin=c]{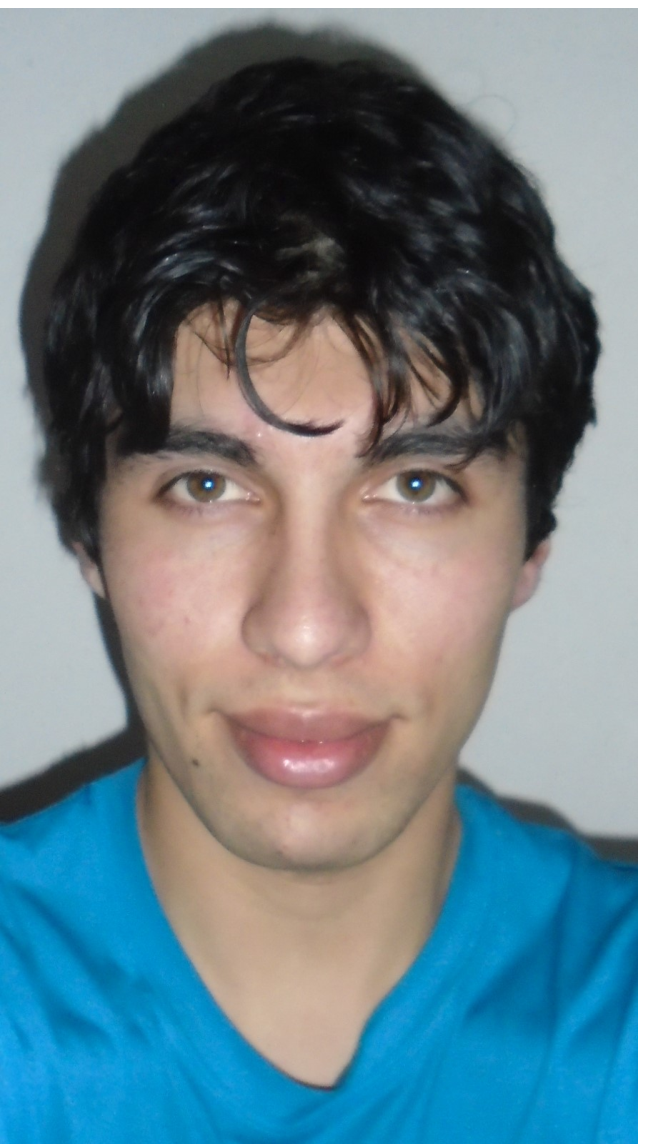}}
\end{minipage}
\hfill
\begin{minipage}{0.8\textwidth}
\begin{small}
Ricardo Heras is an undergraduate student in Astrophysics at University College London. He has been inspired by Feynman's teaching philosophy that if one cannot provide an explanation for a topic at the undergraduate level then it means one doesn't really understand this topic. His interest in understanding physics has led him to publish several papers in \emph{The European Journal of Physics} on the teaching of electromagnetism and special relativity. He has also authored research papers on magnetic monopoles, pulsar astrophysics, history of relativity, and two essays in Physics Today. For Ricardo the endeavour of publishing papers in physics represents the first step towards becoming a physicist driven by ``The pleasure of finding things out.''
\end{small}
\end{minipage}

\appendix

\section{Derivation of Equations (3) and (12)}
\label{A}

\noindent The curl of Equation (2) gives
\begin{align}
\gradv \times \v A_L &=\gradv \times \bigg( \gradv\times \bigg\{ \int_L \frac{g \,d\v l'}{|\v x-\v x'|} \bigg\}\bigg)\nonumber\\
\nonumber &=\, \gradv\bigg(\!\gradv\! \cdot\! \bigg\{ \! \int_L \!\frac{g \,d\v l'}{|\v x\!-\!\v x'|} \bigg\} \!\bigg)\!-\!\gradv^2 \bigg\{ \!\int_L \!\frac{g \,d\v l'}{|\v x\!-\!\v x'|}  \!\bigg\}
\\ & = \,  g\gradv \!\!\int_L \! \gradv \!\cdot\! \bigg(\!\frac{d\v l'}{|\v x\!-\!\v x'|}\!\bigg) \!-\! g\! \int_L \! \gradv^2\bigg(\!\frac{1}{|\v x\!-\!\v x'|}\!\bigg)d\v l'.
\end{align}
Using the result $\gradv \cdot (d \v l'/|\v x\!-\!\v x'|) = d\v l' \cdot \gradv(1/|\v x\!-\!\v x'|),$ the first integral becomes
\begin{align}
\nonumber \int_L  \gradv \cdot \bigg(\frac{d\v l'}{|\v x\!-\!\v x'|}\bigg)=& \int_L  \gradv\bigg(\frac{1}{|\v x\!-\!\v x'|}\bigg)\cdot d\v l'
\\ \nonumber = & - \int_L  \gradv'\bigg(\frac{1}{|\v x\!-\!\v x'|}\bigg)\cdot d\v l'
\\  = &-\frac{1}{|\v x\!-\!\v x'|}.
\end{align}
Considering Equation (A2), the first term of Equation (A1) yields the field of the magnetic monopole
\begin{align}
g\gradv \int_L  \gradv \cdot \bigg(\frac{d\v l'}{|\v x\!-\!\v x'|}\bigg) = g \gradv\bigg( \!\!-\frac{1}{|\v x\!-\!\v x'|} \bigg)= \frac{g}{R^2}\hat{\v R},
\end{align}
where we have used $\gradv (1/|\v x- \v x'|)= -\hat{\v R}/R^2.$ The second term of Equation (A1) yields the magnetic field of the Dirac string
\begin{align}
- g\! \int_L\gradv^2\bigg(\frac{1}{|\v x-\v x'|}\bigg)d\v l'= 4 \pi g\! \int_L \delta(\v x \!-\! \v x')\, d\v l',
\end{align}
where we have used $\gradv^2( 1 / |\v x\!-\!\v x'|)=-4 \pi \delta(\v x \!-\! \v x').$ The Addition of Equations (A3) and (A4) yields Equation (3).

To derive Equation (12), we first take the curl of Equation (11),
\begin{align}
\nonumber \gradv \times \v A_L =&\gradv \times \bigg(\gradv \times \bigg\{\uvec{z}  \int\limits_{-\infty}^{0}\frac{g\, dz'}{|\v x\!-\!z'\hat{\v z}|}\bigg\}\bigg)
\\ \nonumber=&\, \gradv\bigg(\!\gradv\! \cdot\! \bigg\{\hat{\v z} \!\! \int\limits_{-\infty}^{0}\!\!\frac{g \,dz'}{|\v x\!-\!z'\uvec{z}|}\bigg\} \bigg)\!-\!\gradv^2 \bigg\{\!\uvec{z} \!\! \int\limits_{-\infty}^{0}\!\!\frac{g \,dz'}{|\v x\!-\!z'\uvec{z}|}\bigg\}
\\= & \,  g\gradv\!\!\!\int\limits_{-\infty}^{0}\!\! \frac{\partial}{\partial z}\bigg(\! \frac{dz'}{|\v x\!-\!z'\uvec{z}|} \!\bigg)\!-\! g\,\hat{\v z}  \!\!\!\int\limits_{-\infty}^{0}\!\!\gradv^2\bigg(\frac{ dz'}{|\v x\!-\!z'\hat{\v z}|}\bigg).
\end{align}
To simplify the first term we may write
\begin{align}
\frac{\partial}{\partial z}\bigg( \frac{1}{|\v x\!-\!z'\uvec{z}|} \bigg)= - \frac{z-z'}{\big(x^2+y^2+(z-z')^2\big)^{3/2}},
\end{align}
so that
\begin{align}
\int\limits_{-\infty}^{0}\!\!\frac{\partial}{\partial z} \bigg(\frac{d z' }{|\v x\!-\!z'\uvec{z}|}\bigg)= -\!\!\int\limits_{-\infty}^{0}\! \frac{z-z'}{\big(x^2+y^2+(z-z')^2\big)^{3/2}} \,dz'.
\end{align}
Consider the substitution $u(z')=x^2+y^2+(z-z')^2$. Hence, $du = -2(z-z')dz',$ and the right-hand side of the integral in Equation (A7) takes the form
\begin{align}
\frac{1}{2}\lim_{\beta\to\infty}\int_{u(z'=-\beta)}^{u(z'=0)}\!\frac{ du }{u^{3/2}} =  \lim_{\beta\to\infty} \frac{-1}{\sqrt{u}}\bigg|^{u(z'=0)}_{u(z'=-\beta)}
= -\frac{1}{|\v x|}+ \lim_{z'\to-\infty} \frac{1}{|\v x - z'\uvec{z}|}= -\frac{1}{r}.
\end{align}
Using this result in the first term in Equation (A5) we obtain the monopole field
\begin{align}
g\gradv\!\!\!\int\limits_{-\infty}^{0}\!\! \frac{\partial}{\partial z}\bigg(\! \frac{dz'}{|\v x\!-\!z'\uvec{z}|} \!\bigg)=g \gradv \bigg(\!\!-\frac{1}{r} \bigg)=  \frac{g}{r^2}\hat{\v r}.
\end{align}
To simplify the second term in Equation (A5) consider
\begin{align}
\nonumber \gradv^2\bigg(\frac{1}{|\v x\!-\!z'\uvec{z}|}\bigg)\!=&-4\pi \delta(\v x\!-\!z'\uvec{z})\!
\\=&-4 \pi\delta(x)\delta(y)\delta(z\!-\!z').
\end{align}
Using this equation in the second term of Equation (A5) we obtain the string field
\begin{align}
\nonumber - g\hat{\v z}\!\!\!\int\limits_{-\infty}^{0}\!\!\gradv^2\bigg(\!\frac{d z'}{|\v x\!-\!z'\uvec{z}|}\!\bigg)=&4\pi g\delta(x)\delta(y)\bigg\{\!\!\int\limits_{-\infty}^{0}\!\delta(z\!-\!z')dz'\bigg\}\hat{\v z}
\\
=&4 \pi g \delta(x)\delta(y)\Theta(-z)\hat{\v z},
\end{align}
where in the last step we have used the integral representation of the step function $\Theta(\xi\!-\!\alpha) =\int_{- \infty}^{\xi} \delta(\tau\!-\!\alpha)d\tau$ to identify the quantity within the brackets $\{\,\,\,\}$ in Equation (A11) as $\Theta(-z)=\int_{-\infty}^{0}\delta(z\!-\!z')dz'.$ Addition of Equations (A9) and (A11) yields Equation (12).

\section{Derivation of Equation (13)}
\label{B}
\noindent Using Equation (11), we obtain
\begin{align}
\nonumber \v A_L=&\,g \gradv \times \hat{\v z}\int\limits_{-\infty}^{0}\frac{ dz'}{|\v x-z'\hat{\v z}|}
 = \, g\bigg( \frac{\partial}{\partial y}\hat{\v x}  - \frac{\partial}{\partial x} \hat{\v y}\bigg)\int\limits_{-\infty}^{0} \frac{dz'}{|\v x-z'\hat{\v z}|}
\\ =& \,g\!\!\int\limits_{-\infty}^{0}\! \bigg\{\frac{\partial}{\partial y}\bigg(\frac{\hat{\v x}}{|\v x\!-\!z'\hat{\v z}|}\bigg)-\frac{\partial}{\partial x}\bigg(\frac{\hat{\v y}}{|\v x\!-\!z'\hat{\v z}|}\bigg)\bigg\} \,dz'.
\end{align}
Now,
\begin{align}
\frac{\partial}{\partial y}\bigg(\frac{1}{|\v x-z'\hat{\v z}|}\bigg) =& -\frac{y}{(x^2+y^2 +(z-z')^2)^{3/2}},\\
\frac{\partial}{\partial x}\bigg(\frac{1}{|\v x-z'\hat{\v z}|}\bigg) =& -\frac{x}{(x^2+y^2 +(z-z')^2)^{3/2}}.
\end{align}
Inserting these equations in Equation (B1) we obtain
\begin{align}
 \v A_L=&g \big(\!- y\hat{\v x} + x\hat{\v y}\big)\int\limits_{-\infty}^{0}\frac{dz'}{(x^2+y^2 +(z-z')^2)^{3/2}}.
\end{align}
The integral can be solved by a variable change and an appropriate substitution. We can write $(z-z')^2=(z'-z)^2.$ Now we let $u(z') = z'-z$ so that $du=dz'.$ Hence, the integral in Equation (B4) may be written as
\begin{align}
\lim_{\beta\to\infty}\int_{u(z'=-\beta)}^{u(z'=0)}&\frac{du}{(x^2+y^2 +u^2)^{3/2}}.
\end{align}
An appropriate substitution for solving this integral is  $u(v) = \sqrt{x^2 +y^2} \tan(v),$ where $v=\tan^{-1}( u/\sqrt{x^2+y^2}).$ This relation assumes $\sqrt{x^2 +y^2}\neq 0,$ indicating that the negative $z$-axis associated to the Dirac string has been avoided. It follows that $du=\sec^2(v)dv$ and then the integral in Equation (B5) becomes
\begin{align}
\lim_{\beta\to\infty}\int_{v(u(z'=-\beta))}^{v(u(z'=0))}\frac{\sqrt{x^2+y^2} \sec^2(v)}{\big((x^2+y^2)(\tan^2(v)+1)\big)^{3/2}}\,dv.
\end{align}
Using the identity $\sec^2(v)=\tan^2(v) +1$, the denominator in Equation (B6) simplifies to $(x^2+y^2)^{3/2}\sec^3(v).$ It follows
\begin{align}
\nonumber \frac{1}{x^2+y^2}\lim_{\beta\to\infty}\int_{v(u(z'=-\beta))}^{v(u(z'=0))}\frac{dv}{\sec(v)} =& \frac{1}{x^2\!+\!y^2}\lim_{\beta\to\infty}\!\int_{v(u(z'=-\beta))}^{v(u(z'=0))} \!\cos(v)\,dv\\ =& \lim_{\beta\to\infty}\frac{\sin(v)}{x^2+y^2}\bigg|^{v(u(z'=0))}_{v(u(z'=-\beta))},
\end{align}
where $\cos(v) = 1/\sec(v)$ has been used. Considering the identity $\sin\big(\tan^{-1}(\alpha)\big)=\alpha/\sqrt{\alpha^2+1}$, we can easily evaluate Equation (B7)
\begin{align}
\nonumber \lim_{\beta\to\infty}\,\frac{\sin(v)}{x^2+y^2}\bigg|^{v(u(z'=0))}_{v(u(z'=-\beta))}
=&\, \bigg(\!\frac{1}{x^2+y^2}\!\bigg)\lim_{\beta\to\infty} \frac{u}{\sqrt{x^2+y^2}\sqrt{\frac{u^2}{x^2+y^2} +1}}\bigg|^{u(z'=0)}_{u(z'=-\beta)}
\\ \nonumber=&\,\bigg(\!\frac{1}{x^2+y^2}\!\bigg) \lim_{\beta\to\infty}\frac{z'-z}{\sqrt{x^2+y^2+(z\!-\!z')^2}}\bigg|^{z'=0}_{z'=-\beta}
\\ =&\,\frac{1}{x^2+y^2}\bigg(1 -\frac{z}{\sqrt{x^2+y^2+z^2}}\bigg).
\end{align}
From Equation (B8) in Equation (B4) we obtain
\begin{align}
\v A_L=\,g\frac{\big(\!- y\hat{\v x} + x\hat{\v y}\big)}{x^2+y^2}\bigg(1 -\frac{z}{\sqrt{x^2+y^2+z^2}}\bigg).
\end{align}
Considering spherical coordinates $r=\sqrt{x^2+y^2+z^2},$ $r\sin\theta=\sqrt{x^2+y^2},$ $r\cos\theta=z$ and $\hat{\phi}=(- y\hat{\v x}+x\hat{\v y})/(\sqrt{x^2+y^2})$, Equation (B9) takes the form $\v A_L= g [(1 - \cos\theta)/(r \sin\theta)]\hat{\phi},$ which is Equation (13).

\section{Derivation of Equation (14)}
\label{C}
\noindent Consider the first equality in Equation (14)
\begin{align}
\v A_{L'}-& \v A_L = \,g\gradv\times \oint_C \frac{d\v l'}{|\v x-\v x'|}.
\end{align}
Using Stoke's theorem and $\gradv(1/|\v x \!-\! \v x'|)=-\gradv'(1/|\v x \!-\!\v x'|),$ Equation (C1) becomes
\begin{align}
\nonumber \v A_{L'} - \v A_L =& -g \gradv \times \int_{S} \gradv'\bigg(\frac{1}{|\v x \!-\! \v x'|} \bigg)\times d\v a'
\\ \nonumber = & \, \gradv \times \bigg( \gradv \times \bigg\{\int_{S} \frac{g \,d\v a'}{|\v x \!-\! \v x'|}\bigg\} \bigg)
\\ =&\, \gradv\bigg(\!\gradv\! \cdot\! \bigg\{\!\int_{S} \frac{g \,d\v a'}{|\v x \!-\! \v x'|}\bigg\} \bigg)\!-\!\gradv^2 \bigg\{ \!\int_{S} \frac{g \,d\v a'}{|\v x\!-\!\v x'|} \bigg\}.
\end{align}
Making use of $\gradv \cdot (d \v a'/|\v x\!-\!\v x'|) =d\v a' \cdot \gradv(1/|\v x\!-\!\v x'|)$  Equation (C2) reads
\begin{align}
\nonumber \v A_{L'}\!-\! \v A_L =& \,g \gradv \!\!\int_{S} \!\gradv \bigg(\!\frac{1}{|\v x \!-\! \v x'|}\!\bigg) \!\cdot\! d \v a' \!-\! g \!\!\int_{S}\!\gradv^2 \bigg(\!\frac{1}{|\v x \!-\! \v x'|}\!\bigg)d\v a'
\\ = & \,g \gradv \!\!\int_{S} \frac{(\v x' \!-\!\v x)\cdot d \v a'}{|\v x \!-\! \v x'|^3}+ 4\pi g \!\int_{S} \! \delta(\v x \!-\!\v x')\,d\v a',
\end{align}
where we have used $\gradv(1/|\v x\!-\!\v x'|) = -(\v x \!-\!\v x')/|\v x\!-\!\v x '|^3$ and $\gradv^2(1/|\v x\!-\!\v x'|)\!=\!-4 \pi \delta(\v x\!-\!\v x').$ The integral in the first term of Equation (C3) is the solid angle \cite{75}
\begin{align}
\Omega(\v x) = \int_{S} \frac{(\v x' \!-\!\v x)\cdot d \v a'}{|\v x \!-\! \v x'|^3},
\end{align}
and therefore
\begin{align}
\v A_{L'}- \v A_L = \,g \gradv\Omega + 4\pi g \!\int_{S} \!\delta(\v x \!-\!\v x')\,d\v a'.
\end{align}
The delta integral  contribution vanishes at any point $\v x$ not on the surface $S$ and can therefore be dropped \cite{7}. Thus we obtain $\v A_{L'}- \v A_L = \,g \gradv\Omega,$ which is Equation (14). Discussions on Equation (C5) can be found in References \cite{7,12,84}.

\section{Derivation of Equation (40)}
\label{D}
\noindent Consider the first vector potential given in Equation (32), namely $\v A' = [g(1\!-\! \cos\theta)/(r\sin\theta)]\hat{\phi}$ which is valid for $z<0.$ For convenience, we express this potential in cylindrical coordinates
\begin{align}
\v A' = \frac{g}{\rho}\bigg(1- \frac{z}{\sqrt{\rho^2+z^2}}  \bigg)\hat{\phi}.
\end{align}
where we have used $\cos \theta= z/\sqrt{\rho^2 +z^2},$ and $r\sin \theta= \rho,$ with $\rho=\sqrt{x^2 + y^2}.$ A regularised form of this potential can be obtained by making the replacements \cite{58}: $1/\rho\rightarrow\Theta(\rho-\varepsilon)/\rho,$ and $z/\sqrt{\rho^2 +z^2}\rightarrow z/\sqrt{\rho^2 +z^2+\varepsilon^2},$ where $\Theta$ is the step function and $\varepsilon>0$ is an infinitesimal quantity. It follows
\begin{align}
\v A'_{\varepsilon} = \frac{g\,\Theta(\rho-\varepsilon)}{\rho}\bigg(1 - \frac{z}{\sqrt{\rho^2+z^2+\varepsilon^2}}  \bigg)\hat{\phi}.
\end{align}
Clearly, in the limit $\varepsilon \to 0$ we recover Equation (D1). Consider now the definition of the curl of the generic vector $\bfF= \bfF[0, F_\phi(\rho,z),0]$ in cylindrical coordinates given in Equation (61). Using this definition in Equation (D2) we obtain
\begin{align}
\nonumber\gradv \times \v A'_{\varepsilon}=& -\frac{g \Theta (\rho-\varepsilon)}{\rho} \bigg( \frac{\rho^2 + \varepsilon^2}{(\rho^2+z^2+\varepsilon^2)^{3/2}} \bigg)\hat{\rho}
+ \frac{g\Theta(\rho - \varepsilon)}{\rho} \bigg( \frac{z}{(\rho^2 +z^2 +\varepsilon^2)^{3/2}} \bigg)\hat{\v z}
\\ \nonumber  &  + \bigg\{\frac{g \delta(\rho - \varepsilon)}{\rho}- \frac{gz\delta(\rho-\varepsilon)}{\rho \sqrt{\rho^2+z^2+\varepsilon^2}}\bigg\}\hat{\v z}
\\ \nonumber =\,& \frac{g \Theta(\rho-\varepsilon)}{(\rho^2 +z^2 + \varepsilon^2)}\bigg( \frac{\rho \hat{\rho}+z \hat{\v z}}{\sqrt{\rho^2+z^2 +\varepsilon^2}} \bigg) -  \frac{\varepsilon^2 \,g \Theta (\rho\!-\!\varepsilon)\hat{\rho}}{\rho(\rho^2+z^2+\varepsilon^2)^{3/2}}
\\  & + \bigg\{ \frac{g \delta(\rho-\varepsilon)}{\rho} - \frac{g z \delta(\rho-\varepsilon)}{\rho \sqrt{\rho^2+z^2+\varepsilon^2}} \bigg\}\hat{\v z}.
\end{align}
In the last term enclosed within the brackets $\{\,\,\,\},$ we add the exact zero quantity $\big[g\delta(\rho-\varepsilon)/\rho- g\delta(\rho-\varepsilon)/\rho\big]\hat{\v z} \equiv 0,$ and obtain
\begin{align}
\nonumber\gradv \times \v A'_{\varepsilon}= &\, \frac{g\, \Theta(\rho-\varepsilon)}{(\rho^2 +z^2 + \varepsilon^2)}\bigg( \frac{\rho \hat{\rho}+z \hat{\v z}}{\sqrt{\rho^2+z^2 +\varepsilon^2}} \bigg)+ \frac{2g\, \delta(\rho -\varepsilon)\hat{\v z}}{\rho}
\\ & - \frac{\varepsilon^2 \,g \Theta (\rho\!-\!\varepsilon)\hat{\rho}}{\rho(\rho^2+z^2+\varepsilon^2)^{3/2}} -\frac{g \,\delta(\rho-\varepsilon)}{\rho}\bigg( \frac{\sqrt{\rho^2 +z^2 + \varepsilon^2}+z}{\sqrt{\rho^2 + z^2 +\varepsilon^2}} \bigg)\hat{\v z}.
\end{align}
This is a regularised form of the magnetic field produced by the potential $\v A'_{\varepsilon}.$ The first two terms of Equation (D4) are the only non-vanishing terms in the limit $\varepsilon \rightarrow 0.$ The third term is shown to vanish easily because there is a term $\varepsilon^2$ in the numerator. However, it is not clear why the last term should vanish. Let us analyse this term. Consider an arbitrary point $z_0$ on the negative $z$-axis. For small $\varepsilon,$ we can make the replacement \cite{22}:
$\sqrt{\rho^2 +z^2 + \varepsilon^2}+z \to (\rho^2+\varepsilon^2)/(2z_0).$
With this replacement, the last term in Equation (D4) becomes
\begin{align}
\bigg(\frac{g \,\delta(\rho-\varepsilon)\rho}{2 z_0\sqrt{\rho^2 + z^2 +\varepsilon^2}} + \frac{g \,\delta(\rho-\varepsilon)\,\varepsilon^2}{2\rho z_0\sqrt{\rho^2 + z^2 +\varepsilon^2}}  \bigg) \hat{\v z}.
\end{align}
In the limit $\varepsilon \rightarrow 0,$ it follows that Equation (D5) vanishes because $\varepsilon^2 \rightarrow 0$ and $\delta(\rho) \rho =0.$ Hence,
\begin{align}
\nonumber\lim_{\varepsilon\to 0} \gradv \times \v A'_{\varepsilon} =& \lim_{\varepsilon\to 0}\bigg\{\frac{g \Theta(\rho-\varepsilon)}{(\rho^2 +z^2 + \varepsilon^2)}\bigg( \frac{\rho \hat{\rho}+z \hat{\v z}}{\sqrt{\rho^2+z^2+ \varepsilon^2}} \bigg) + \frac{2g \delta(\rho-\varepsilon)\hat{\v z}}{\rho}\bigg\}
\\ = &\; g\frac{\hat{\v r}}{r^2} + 4 \pi g \delta(x)\delta(y)\Theta(-z)\hat{\v z},
\end{align}
where we have used $\hat{\v r}=(\rho \hat{\rho} +z\hat{\v z}) /(\sqrt{\rho^2 +z^2}),$  and inserted $\Theta(-z)=1$ to specify that this expression is valid only for $z<0.$

\section{Derivation of Equation (89)}
\label{E}
\noindent Consider the electromagnetic angular momentum of the Thomson dipole whose configuration is shown in Fig.~\ref{Fig10}. The electric and magnetic fields of this dipole are
\begin{align}
\v E = q\,\frac{(\v x + \v a /2)}{|\v x + \v a /2|^3}, \quad \v B = g\,\frac{(\v x - \v a /2)}{|\v x - \v a /2|^3}.
\end{align}
These fields satisfy
\begin{align}
\gradv\cdot \v E =&\, 4 \pi q \delta(\v x+\v a/2), \quad\gradv \times \v E = 0,
\\ \gradv \cdot \v B =&\, 4 \pi g \delta(\v x-\v a/2), \quad \gradv \times \v B = 0.
\end{align}
In particular, the electric field can be expressed as the gradient of the electric potential $\v E = - \gradv \Phi,$ where
\begin{align}
\Phi(\v x) = \frac{q}{|\v x + \v a /2|}.
\end{align}
Using $\v E = - \gradv \Phi,$  we write $\v E \times \v B = - \gradv \Phi \times \v B,$ which combines with  $\gradv \times (\Phi \v B) =\Phi\gradv \times \v B+\gradv \Phi \times \v B $ to obtain $\v E \times \v B = - \gradv \times (\Phi \v B).$ If we define the vector $\v W=\Phi\v B,$ then  $\v E \times \v B = - \gradv \times \v W.$ Using this expression in the integrand of Equation (88), we obtain
\begin{align}
\v x \times (\v E \times \v B)= - \v x \times (\gradv \times \v W).
\end{align}
To write Equation (E5) in an appropriate form, we can use the following identity expressed in index notation \cite{61}:
\begin{align}
\big[\v x \times \!\big(\gradv \!\times \!\v W\big)\big]^i = &\, -\partial_j\big(x^jW^i-2W^jx^i\big)+\partial^i\big( x_jW^j\big) -2 x^i\partial_jW^j.
\end{align}
Here summation convention on repeated indices is adopted and $\varepsilon^{ijk}$ is the Levi-Civita symbol with $\varepsilon^{123}=1$ and $\delta^{i}_{j}$ is the Kronecker delta. Equation (E6) can be readily verified. First we write
\begin{align}
\nonumber \big[\v x \times \!\big(\gradv \!\times \!\v W\big)\big]^i=& \, \varepsilon^{ijk}x_j \big( \gradv \times \v W\big)_k
\\ \nonumber =&\,\varepsilon^{ijk} x_j \varepsilon_{klm} \partial^l W^m
\\ \nonumber =& \, (\delta^{i}_{l} \delta^{j}_{m}-\delta^{j}_{l} \delta^{i}_{m}) \, x_j \partial^l W^m
\\  =& \,  x_m \partial^i W^m - (x_m\partial^m) W^i,
\end{align}
where we have used the identity $\varepsilon^{ijk}\varepsilon_{klm}=\delta^{i}_{l} \delta^{j}_{m}-\delta^{j}_{l} \delta^{i}_{m}.$
Now, consider the identically zero quantities
\begin{align}
2\big( \partial_m W^m x^i - \partial_m W^m x^i \big)\equiv 0,
\\
\big(\partial^i x_m W^m + 2 W^m\partial_m x^i - \partial_m x^m W^i  \big)\equiv 0.
\end{align}
Adding Equations (E8) and (E9) to Equation (E7), we obtain Equation (E6). When Equation (E6) is integrated over a volume, the first two terms of the right-hand side can be transformed into surface integrals which are shown to vanish for a large $r.$ Therefore,
\begin{align}
\nonumber \int_{V}\big[\v x \times \big(\v E \times  \v B \big)\big]^i \,d^3x=&\, 2\int_{V} x^i\partial_jW^j\,d^3x =\,2 \int_{V}x^i(\partial_j\Phi B^j+\Phi\partial_jB^j)\,d^3x
\\
=&-2\!\int_{V} \!x^i(E_j B^j\,)\,d^3x +2\!\int_{V}\!x^i\Phi(\partial_jB^j)\,d^3x.
\end{align}
Using Equation (E10) in Equation (88), we obtain
\begin{align}
\v L_\texttt{EM}=-\frac{1}{2\pi c}\!\int_{V} \v x \,\big(\v E\!\cdot\!\v B\big)\,d^3x + \frac{1}{2\pi c}\!\int_{V}\v x\, \Phi (\gradv\!\cdot\!\v B)\,d^3x = \,\frac{1}{2\pi c}\!\int_{V}\v x\, \Phi (\gradv\!\cdot\!\v B)\,d^3x,
 \end{align}
where the integral in the first term has vanished because integrand is an odd function of $\v x$ for the chosen origin. Using Equations (E3) and (E4), we substitute $\gradv\cdot\v B=4\pi g\delta(\v x-\v a/2)$ and $ \Phi=q/|\v x + \v a /2|$ into the second integral, obtaining the expected result
\begin{align}
\v L_\texttt{EM}=\, \frac{2qg}{c}\!\int_{V} \delta(\v x -\v a/2) \bigg(\frac{\v x}{|\v x + \v a /2|}\bigg)d^3x=\, \frac{2qg}{c}\,\, \frac{\v x}{|\v x + \v a /2|}\bigg|_{\v x = \v a /2}=\frac{qg}{c}\hat{\v a}.
\end{align}

\end{document}